%% file: main.tex
\documentclass[lettersize,journal]{IEEEtran}

\usepackage{amssymb}
\usepackage{amsmath,amsfonts}
\usepackage{mathtools} 
\usepackage[ruled,vlined,linesnumbered]{algorithm2e}
\usepackage{parskip}
\usepackage{graphicx}
\usepackage{textcomp}
\usepackage{xcolor}
\usepackage{comment}
\usepackage[frozencache,cachedir=.]{minted}
\usepackage{multirow}
\usepackage{multicol}
\usepackage{listings}
\usepackage{circledsteps}
\usepackage{parcolumns}
\usepackage{stfloats}
\usepackage{color,soul}
\usepackage{xurl}
\usepackage{threeparttable}
\usepackage{tikz}
\usepackage{xspace}
\usepackage{cprotect}
\usepackage{verbatimbox}
\usepackage{caption}
\usepackage{subcaption}
\usetikzlibrary{calc}
\usepackage{booktabs}
\usepackage{threeparttable}
\usepackage{circledsteps}
\usepackage{pzccal}
\usepackage{siunitx}
\usepackage{hyperref}
\usetikzlibrary{arrows,automata,positioning}

\definecolor{codegreen}{rgb}{0,0.6,0}
\definecolor{codegray}{rgb}{0.5,0.5,0.5}
\definecolor{codepurple}{rgb}{0.58,0,0.82}
\definecolor{backcolour}{rgb}{0.95,0.95,0.92}
\definecolor{darkblue}{rgb}{0.0, 0.0, 0.55}
\definecolor{darkgreen}{rgb}{0.0, 0.2, 0.13}
\definecolor{darkred}{rgb}{0.6, 0.0, 0.0}
\definecolor{darkorange}{rgb}{1.0, 0.27, 0.0}

\lstdefinestyle{mystyle}{
    backgroundcolor=\color{backcolour},   
    commentstyle=\color{codegreen},
    keywordstyle=\color{magenta},
    numberstyle=\tiny\color{codegray},
    stringstyle=\color{codepurple},
    breakatwhitespace=false,         
    breaklines=true,                 
    captionpos=b,                    
    numbers=left,                    
    showspaces=false,                
    showstringspaces=false,
    showtabs=true,                  
}

\makeatletter
\def\namedlabel#1#2{\begingroup
    #2%
    \def\@currentlabel{#2}%
    \phantomsection\label{#1}\endgroup
}
\makeatother

\def\BibTeX{{\rm B\kern-.05em{\sc i\kern-.025em b}\kern-.08em
    T\kern-.1667em\lower.7ex\hbox{E}\kern-.125emX}}
    
\usepackage{tcolorbox}
\newtcolorbox{mtbox}[1]{left=0.25mm, right=0.25mm, top=0.25mm, bottom=0.25mm, colframe=red!50!black, boxrule=0.5pt, title={#1}, fonttitle=\bfseries, coltitle=red!50!black, attach title to upper={\ --\ }}

\usepackage[framemethod=tikz]{mdframed}
\mdfdefinestyle{remarkstyle}{
backgroundcolor=lightgray,
innerleftmargin=2pt,
innerrightmargin=2pt,
innertopmargin=2pt,
innerbottommargin=2pt,
linewidth=0.5pt,
}

\newcounter{summ}[section]

\newcommand{\ziming}[1]{%
  \begingroup
  \definecolor{hlcolor}{RGB}{20, 255, 20}\sethlcolor{hlcolor}%
  \textcolor{black}{\hl{\textit{\textbf{Ziming:} #1}}}%
  \endgroup
}

\pgfkeys{/csteps/inner xsep=1.5pt}
\pgfkeys{/csteps/inner ysep=1.5pt}
\pgfkeys{/csteps/inner color=white} 
\pgfkeys{/csteps/outer color=black} 
\pgfkeys{/csteps/fill color=black}  

\newcommand{\solution}[1]{%
	\begingroup
	\definecolor{hlcolor}{RGB}{20, 255, 255}\sethlcolor{hlcolor}%
	\textcolor{black}{\hl{\textit{\textbf{Solution:} #1}}}%
	\endgroup
}

\newcommand{\clfat}[1]{%
	\begingroup
	\definecolor{hlcolor}{RGB}{255, 255, 153}\sethlcolor{hlcolor}%
	\textcolor{black}{\hl{\textit{\textbf{C-FLAT:} #1}}}%
	\endgroup
}
\newcommand{\qes}[1]{%
	\begingroup
	\definecolor{hlcolor}{RGB}{255, 204, 204}\sethlcolor{hlcolor}%
	\textcolor{black}{\hl{\textit{\textbf{Question:} #1}}}%
	\endgroup
}
\newcommand{\tomal}[1]{%
	\begingroup
	\definecolor{hlcolor}{RGB}{233, 249, 12}\sethlcolor{hlcolor}%
	\textcolor{black}{\hl{\textit{\textbf{Tomal:} #1}}}%
	\endgroup
}

\newcommand{\enola}{{\scshape Enola}\xspace}

\author{
\begin{tabular}[t]{ccc}
	\textnormal{Md Armanuzzaman} & \textnormal{Engin Kirda} & \textnormal{Ziming Zhao} \\ 
	\textit{marmanuzzaman@utep.edu} & \textit{ek@ccs.neu.edu} & \textit{z.zhao@northeastern.edu}  \\
	CactiLab, University of Texas at El Paso & Northeastern University & CactiLab, Northeastern University 
\end{tabular}
}

\begin{document}
	
\title{\enola: Linear-Space and Low-Overhead Control-Flow Attestation for Microcontroller-based Systems}
\maketitle

\begin{abstract}
Control-Flow Attestation (CFA) aims to precisely verify execution paths to a remote verifier. However, existing solutions are fundamentally incompatible with resource-constrained, microcontroller-based systems due to the following reasons:  
(1) the overhead of transmitting measurement and trace data scales poorly (i.e., often exponentially) with the number of basic blocks or linearly with the length of execution traces, making such approaches impractical even on high-end microprocessor-based systems and entirely infeasible on microcontrollers;  
(2) cryptographic keys and measurement data are typically stored in memory, rendering them vulnerable to cold boot and memory corruption attacks, which is a serious threat for field-deployed devices; and  
(3) reliance on software-based measurements, combined with frequent context switches between the Rich Execution Environment (REE) and the Trusted Execution Environment (TEE), introduces significant performance overhead.

In this paper, we present \enola, a lin\ul{e}ar-space and low-overhead co\ul{n}tr\ul{o}l-f\ul{l}ow \ul{a}ttestation solution for microcontroller-based systems.
\enola achieves linear transmission complexity with basic blocks, guaranteeing its scalability for larger programs. 
Moreover, \enola uses hardware-assisted measurement computation present in
off-the-shelf devices, and avoids key storage in memory. 
\enola also allocates general-purpose registers for measurements to thwart memory corruption attacks. 
\enola significantly reduces context switching overhead by eliminating transitions from REE to TEE for backward-edge measurements.
We developed the \enola compiler using LLVM passes and a custom attestation
engine targeting the ARMv8.1-M architecture. Our evaluation shows that
\enola reduces data transmission overhead by an average of 52x on the Embench, while maintaining performance comparable to (or exceeding) that of existing approaches. Furthermore, \enola is validated on large, real-world applications such as wolfSSL, enabling scalable control-flow attestation on microcontrollers for the first time.
\end{abstract}

\maketitle

\input{introduction}

\input{problem}

\input{background}


\input{approach}

\input{analysis}

\input{implementation}

\input{evaluation}
\input{related-work}

\section{Limitations and Future Work}
\textbf{Minimizing runtime}. To reduce runtime overhead and TEE switches, \enola can be realised on top of Blast or leveraging hardware trace components (Micro Trace Buffer to record instruction traces instead of software-based instrumentation.
Furthermore, to eliminate reliance on TEE, unprivileged load/store instructions -- similar to those used in Silhouette~\cite{zhou2020silhouette} and Kage~\cite{du2022kage} -- can be employed to save and access occurrence traces directly in the normal state.
This approach may broaden \enola's applicability while reducing TEE switches, thereby improving overall performance.

\textbf{TOCTOU defense and interrupt}. Defenses like RATA~\cite{de2021toctou} with hardware modifications can be enabled on top of \enola.
\enola currently lacks multitask attestation support for environments like real-time operating systems.
Similarly, its secure interrupt dispatcher is designed for bare-metal applications and does not support nested interrupts, both of which represent promising directions for future enhancement.

\section{Conclusion}
Existing CFA solutions face significant challenges in handling the transmission of large amounts of measurement and/or trace data and often suffer from the slow performance of software-based measurement calculations. These limitations restrict their applicability to small programs or code snippets on high-performance devices.
In this paper, we presented \enola, a novel CFA solution designed specifically for embedded systems. \enola incorporates a novel authenticator with linear space complexity relative to the number of basic blocks and leverages hardware-assisted message authentication code capabilities for efficient measurement computation. It also employs a trusted execution environment and reserves general-purpose registers to mitigate memory corruption attacks.

\section*{Acknowledgments}
This material is based upon work supported in part by the National Science Foundation (NSF) under grants 2508320, 2453496, and 2512972.
Any opinions, findings, conclusions, or recommendations expressed in this material are those of the author(s) and do not necessarily reflect the views of the United States Government or any of its agencies.

\begingroup
\setlength{\itemsep}{0pt}
\setlength{\parskip}{0pt}
\bibliographystyle{IEEEtran}
\bibliography{ref}
\endgroup


\input{bio}

\input{apndx}

\end{document}



\appendix

\section*{Security Analysis}

Our threat model assumes the immutability of normal state code, prevention of code injection, and the hardware security guarantees provided by TrustZone and ARM PA hardware.
However, we consider the presence of a powerful attacker within the normal state of TrustZone, capable of conducting memory corruption, ROP, JOP, and similar attacks.
When exploiting vulnerabilities in the embedded application, the attacker must achieve five key attack prerequisite to circumvent the \enola attestation process: \Circled{1} disable or bypass the instrumented code, \Circled{2} influence \enola measurements $M_f$ and $M_b$, or tamper with measurement key $K_m$ in PA key registers, \Circled{3} execute malicious control flow to generate hash collisions,  \Circled{4} exploit non-secure callable trampoline interfaces in the attestation engine, \Circled{5} replay or corrupt $Auth$, $T_{\mathcal{O}}$, or attestation key $K_a$ to manipulate verification at $\mathbb{V}$.

Prerequisite \Circled{1} is mitigated through the realistic assumption of code immutability, as demonstrated by various previous works~\cite{clements2017protecting,kim2018securing}.
Moreover, any attempt to bypass the instrumentation or interface call will be reflected in the measurements $M_f$ and $M_b$.
For prerequisite \Circled{2}, $M_f$ and $M_b$ measurements are stored in reserved general-purpose registers by the \enola compiler, ensuring that only the instrumented \texttt{pacg} instruction can access these registers.
The compiled binary is also vetted by the \enola \emph{code scanner} to detect any instructions that manipulate the reserved registers or the PA key registers.

An attacker may attempt to launch control-flow violation attacks by exploiting vulnerabilities in the embedded application, but these attacks will be detected during verification through trace and measurements.
To avoid detection, the attacker must construct a sophisticated attack that results in hash collisions (prerequisite \Circled{3}).
However, this is highly infeasible, as the collision probability for the QARMA block cipher with a 64-bit modifier is $2^{-60}$~\cite{avanzi2019qameleon}.
\enola further complicates such attacks by chaining measurements, using the previous control-flow measurement as a modifier for the next one.

TrustZone security guarantees prevent the direct manipulation of $Auth$, $T_{\mathcal{O}}$, or $K_a$, as these are stored and signed in the secure state, while replay attacks are thwarted by the random nonce.
Although $M_f$ and $M_b$ are stored in normal state registers, the failure to achieve prerequisite \Circled{2}, combined with the trusted ARM PA hardware for measurement computation, prevents the corruption of measurement chains in $Auth$.
Consequently, exploiting trampoline interface calls becomes the only viable option to manipulate $Auth$ ($T_{\mathcal{O}}$).
To counter this, the \enola compiler prevents any direct world-switch calls targeting the attestation engine interfaces, allowing only call instructions at the instrumented function locations. 
Additionally, to further safeguard the interfaces from exploitation via indirect calls or jumps, the \enola compiler can adopt commonly used SFI techniques such as address masking for each indirect transfer event.
As a result, attackers are prevented from prerequisite \Circled{4}, which, in combination with TrustZone and ARM PA security guarantees, prevents prerequisite \Circled{5}.

\newpage

\section*{Updated Table 7 with TrustZone Overhead}
We evaluated \enola with TrustZone which resulted in 395.84\%, 565.24\% overhead for O2 and Oz optimizations for Embench applications.
The overhead of wolfSSl application are 124.29\% and 180.04\%.
The two highlighted columns in Table~\ref{t:app} represent the updated runtime overhead after re-evaluation.
\begin{table*}[ht!]
	\centering
	\small
	\resizebox{\textwidth}{!}{
		\begin{tabular}{l|r|r||>{\columncolor{codegray}}r|>{\columncolor{codegray}}r||r||r|r|r}
			\hline
			\multirow{3}{*}{Application} & \multicolumn{4}{c||}{Execution Time (CPU Cycles)}&\multirow{3}{*}{\shortstack{Blast Reported\\ Execution \\Overhead}}   & \multicolumn{2}{c|}{\multirow{2}{*}{ \shortstack{\enola $Auth$ \\ Size (bytes)}}}& \multirow{3}{*}{\shortstack{Blast Reported \\ WPP Size \\ (bytes)}} \\ \cline{2-5}
			& \multicolumn{2}{c||}{Baseline}   & \multicolumn{2}{c||}{\enola }& &\multicolumn{2}{c|}{}&  \\ \cline{2-5} \cline{7-8}
			& \multicolumn{1}{c|}{\texttt{O2}}  & \multicolumn{1}{c||}{\texttt{Oz}} & \multicolumn{1}{c|}{\texttt{O2}}  & \multicolumn{1}{c||}{\texttt{Oz}}&& \multicolumn{1}{c|}{\texttt{O2}}  & \multicolumn{1}{c|}{\texttt{Oz}}& \\ \hline \hline
			\multicolumn{9}{c}{Embench Applications} \\  \hline
			aha-mont64 &72,650,399 & 164,728,199 & 154,480,390 (120.49\%) &494,775,604 (200.35\%)&151\%&168&96&768 \\ \hline
			crc32 &100,156,101 & 102,750,000  &353,131,533 (252.58\%) &353,172,193 (243.72\%)&808\%&48&32&147 \\ \hline
			cubic &2,063,102 &438,423  &2,402,894 (14.77\%)  &582,748 (32.91\%)& 15\%&40& 48&216\\ \hline
			edn &52,169,895 & 87,921,896 &145,771,978 (179.42\%)  &446,972,622 (408.37\%)&75\%&584&264&818 \\ \hline
			huffbench &53,773,298 &66,862,780  &353,093,786 (556.63\%)  &503,017,675 (652.31\%) &86\%&840&256&9,750 \\ \hline
			matmult-int &39,422,817 &73,976,497  &69,330,871 (75.86\%)  &702,371,624 (849.45\%)&32\%&64&80&370 \\ \hline
			md5sum &40,952,437 &56,369,438  &265102190 (547.34\%) &390,440,356 (592.64\%)&-&248&160&- \\ \hline
			minver & 17,995,745&21,936,193  &140,999,106 (683.51\%)  &251,020,965 (1044.32\%)&102\%&360&416&699 \\ \hline
			nbody &5,632,370 &6,776,658  &9,778,662 (73.61\%)  &17,264,580 (154.76\%)& 57\%&192&128&408 \\ \hline
			nettle-aes &81,724,774 &95,756,994  &186332380 (127.99\%)  &236,114,268 (146.57\%)&31\%&408&288&843 \\ \hline
			nettle-sha256 &76,096,054 &89,301,167  &94,027,217 (23.56\%)  &134,224,343 (50.31\%)&9\%&144&144&336 \\ \hline
			nsichneu &77,292,166 & 77,230,494 &739,067,880 (856.20\%)  &742,668,647 (861.626\%)&-&3,056&3,048&- \\ \hline
			primecount &108,246,346 &85,184,383  &1,523,768,160 (1307.68\%)  &1,771,535,786 (1979.64\%)&85\%&512&112&73,478 \\ \hline
			st &9,469,253 & 12,093,362 &10,398,776 (9.81\%)  &25,373,093 (109.81\%)&528\%&160&88&476 \\ \hline
			tarfind &43,636,943 &42,067,145  & 441,261,464 (911.20\%) &437,605,700 (940.26\%)& 518\%&1,072&168&257,756 \\ \hline
			ud &82,104,027 &83,914,248  & 568,849,119 (592.83\%) &735,752,090 (776.79\%)&70\%&608&232&533 \\ \hline 
			\multicolumn{3}{r||}{Average overhead:}  &  \textbf{395.84}\%  & \textbf{565.24}\% &185\%&\multicolumn{3}{c}{}\\ \hline \hline
			
			\multicolumn{9}{c}{wolfSSL Applications} \\  \hline
			AES &  138,517  &148,704 &171,220 (24.50\%)  & 189,945 (27.73\%)&- &336&312&-\\ \hline
			MD5 & 18,957   &19,659& 21,146 (11.55\%) &21,718 (10.47\%)&  - &152&168&-\\ \hline
			HMAC &58,859&70,498&87,504 (48.66\%)&238,318 (238.05\%)&-&488&584&-\\ \hline
			RSA &4,350,974&3,880,909&22,296,595 (412.45\%)& 21,108,491 (443.91\%)&-&1,192&496&-\\ \hline
			\multicolumn{3}{r||}{Average overhead:}  & \textbf{124.29}\%  &\textbf{180.04\%} &-&\multicolumn{2}{c|}{}&-\\ \hline
			
	\end{tabular}}
	\caption{Execution time overhead of baseline and with \enola, $Auth$ size and comparison with Blast. $|Auth|=|T_\mathcal{O}|+8$ (two 32-bit measurements)}
	\label{t:app}
\end{table*}
\bibliographystyle{plain}
\bibliography{./consolidated-reference-bib-file/consolidated-ref}

%% file: introduction.tex
\vspace{-.3cm}
\section{Introduction}

While microcontroller (MCU)-powered embedded systems are ubiquitous, they are vulnerable to cyberattacks, with control-flow hijacking being especially dangerous as it enables arbitrary code execution and full system compromise~\cite{stellios2018survey}.
Numerous efforts have been made to prevent or detect control-flow hijacking in MCU-based systems.
These efforts include forward-edge control-flow integrity~\cite{nyman2017cfi, walls2019control}, shadow stacks~\cite{zhou2020silhouette, du2022kage}, return address integrity~\cite{almakhdhubmurai2020, kim2023rio}, control-flow violation detection~\cite{tan2023sherloc, wang24insect}, and software-fault isolation.
However, since embedded systems are often deployed in the field, it is crucial not only to detect control-flow hijacking, but also to demonstrate to a remote verifier the runtime control-flow transfers of the system.

Control-Flow Attestation (CFA) was introduced to precisely attest a program's execution path~\cite{ammar2024sok}.
However, existing approaches, such as C-FLAT~\cite{abera2016c}, LO-FAT~\cite{dessouky2017fat}, OAT~\cite{sun2020oat}, and
Blast~\cite{yadav2023whole}, have been designed and evaluated for high-performance microprocessors, limiting their applicability to resource-constrained microcontroller-based systems.
In particular, they face a significant challenge in transmitting large measurement and/or trace data.
The size of these transmissions grows either exponentially with the number of basic blocks in the program, or linearly with the runtime execution trace.
As a result, these methods are limited to handling small programs or code snippets.
For instance, C-FLAT generates 1,527 bytes of measurement data during a 1.2-second execution of a \emph{syringe pump} program, while Blast produces 257,756 bytes of trace data for a 38-second execution of the \emph{tarfind} Embench program on Cortex-A53 microprocessors.

Additionally, these approaches store cryptographic keys and measurement results in memory without encryption or integrity protection, rendering them vulnerable to cold boot~\cite{halderman2009lest} or memory corruption attacks. 
This vulnerability persists even within TEEs like ARM TrustZone, which provide memory isolation, but do not encrypt memory.
Moreover, many existing CFA solutions rely on software-based keyed hash functions for measurement computation, incurring substantial latency on resource-constrained MCUs.
This overhead is compounded by frequent context switches from REE to TEE for both forward- (call and jump) and backward-edge (return) control-flow events in the attested program.

In this paper, we present \enola, a lin\ul{e}ar-space and low-overhead co\ul{n}tr\ul{o}l-f\ul{l}ow \ul{a}ttestation solution for microcontroller-based embedded systems. 
The primary objectives and challenges
of \enola are: (1) to algorithmically minimize the footprint of trace and measurement data, which is architecture-agnostic, (2) to safeguard the measurement and attestation keys and measurement results from memory corruption attacks, and (3) to enable efficient measurement computation on off-the-shelf microcontroller-based devices. Additionally, \enola must remain secure, resist tampering by privileged software, and preserve trace and measurement integrity.

\enola addresses the first challenge by introducing a novel authenticator that combines a
\emph{basic block occurrence trace} with two distinct measurements for the forward and backward execution paths.
This design achieves linear space complexity with respect to the number of basic blocks in the attested program, ensuring scalability to larger programs.
\enola leverages a TEE, such as the Cortex-M TrustZone, to securely acquire the basic block occurrence trace.
Specifically, the \enola compiler instruments the attested programs to report basic block occurrence information to the TEE.

The \enola compiler dedicates two general-purpose registers exclusively for storing forward and backward path measurement chains, preventing memory corruption attacks by ensuring they are not spilled onto memory.
\enola leverages novel hardware-assisted Message Authentication Code (MAC) capabilities, such as the Pointer Authentication (PA) extension in ARMv8.1-M~\cite{liljestrand2019pac} for measurement computations and securely storing keys in PA registers.
Additionally, the cryptographic key registers for PA instructions are initialized within the TEE, and the \enola code scanner verifies that the REE binary is free of instructions, or Return-Oriented Programming (ROP) gadgets capable of altering them.


To address the performance overhead challenge, \enola leverages the MAC hardware capabilities, offering a hundred times speedup over the software-based mechanisms employed by prior solutions.
Although \enola utilizes a TEE, it minimizes context switches from the attested program by eliminating backward path context switches and computing measurements directly in the REE using MAC instructions.

Lastly, compromised Interrupt Service Routines (ISRs) can interfere with authenticator generation by corrupting the attested program state or the reserved measurement registers.
To address this threat, \enola integrates an ISC-FLAT-inspired secure ISR dispatcher with its authenticator realization~\cite{neto2023isc}.
The dispatcher interposes on REE interrupts through the TEE-protected interrupt controller, securely preserves the attested program context and measurement state, restricts ISR access to the program stack, and ensures a controlled return to the interrupted attested program.
Thus, the dispatcher is a supporting protection mechanism for secure authenticator generation rather than the central technical novelty of \enola.

We implemented and evaluated \enola on the Cortex-M85 MCU, using a syringe pump application, Embench, and wolfSSL~\cite{syringe, Embench, wolfSSL} with various optimizations.
Unlike prior solutions, \enola scales to large programs such as wolfSSL while maintaining a competitive performance.
Our evaluations show that \enola drastically reduces the authenticator size: e.g., Blast produces an average of 21,009 bytes of attestation data on Embench, whereas \enola requires only 397 bytes.
The paper contributions are summarized as follows:

\begin{itemize}
\item We note that existing CFA approaches lack a comprehensive analysis of the space complexity associated with trace and measurement data.
To address this gap, we formalize the concept of control-flow attestation, and identify the limitations present in prior studies.
Specifically, we characterize the transmitted-authenticator scalability of representative CFA schemes.

\item We present \enola, 
which introduces a novel authenticator achieving linear space complexity with respect to the number of basic blocks in the attested program.
On the Embench, \enola reduces transmitted data by 52 times on average compared to Blast.
Additionally, to safeguard the authenticator against memory corruption, \enola retains the forward- and backward-edge measurement chains in two compiler-reserved general-purpose registers rather than REE memory, while using TrustZone-protected memory for occurrence trace storage.


\item We implement \enola on an ARMv8.1-M Cortex-M85 microcontroller system leveraging compiler instrumentation, TrustZone, MPU, and PA-based measurement computation.
These mechanisms enable efficient authenticator generation, protect measurement state and measurement keys under the stated threat model, and reduce TEE context switches by computing backward-edge measurements directly in the REE.
Our evaluations demonstrate \enola's effectiveness in substantially reducing transmitted attestation data while achieving performance overhead that is either lower or comparable to existing approaches.
We open-sourced the implementation and evaluation\footnote{https://github.com/CactiLab/ENOLA-Efficient-CFA-for-Embedded-Systems}.
\end{itemize}

%% file: problem.tex
\section{Formalizing the Complexity of CFA}
\label{sec:ps}

\textbf{Modeling control-flow attestation}.
In control-flow attestation, a remote verifier $\mathbb{V}$ requests a prover $\mathbb{P}$ to
execute a program $\mathcal{P}$ by sending a challenge $c$ to ensure the freshness of the request.
$\mathbb{P}$ executes $\mathcal{P}$, and a trusted measurement engine $\mathbb{E}$ within
$\mathbb{P}$ generates an attestation report $\mathcal{R}$, which is then transmitted to
$\mathbb{V}$. The attestation report $\mathcal{R} = (Auth, Sig_{K_a}(Auth, c))$ is composed of a
cumulative authenticator $Auth$ of the control-flow path and a signature over $Auth$ and $c$ using a
key $K_a$.  The authenticator $Auth = (T, M)$ may include a full or partial trace $T$ of the
control-flow path and/or a measurement $M$. 

We model $\mathcal{P}$'s interprocedural Control Flow Graph (CFG) as $G_\mathcal{P} = (V, E)$, where
$V$ represents the set of basic blocks, and $E$ represents the set of control-flow transfers among
the basic blocks defined by $\mathcal{P}$. Each basic block $v_i \in V$ is characterized by an entry
point $v_i.s$, and an exit point $v_i.e$, while each edge $e_i \in E$ is defined by a source address
$e_i.s$ and a destination address $e_i.d$. The full execution trace $T_\mathcal{P}$ of program
$\mathcal{P}$ is represented as a sequence of all non-sequential control-flow transfer destinations
$(e_1.d, \dots, e_l.d)$. We use $n$ to denote the number of forward branch instances in
$T_\mathcal{P}$, which includes conditional jumps and indirect calls/jumps, and $m$ to denote the
number of backward branch instances (i.e., returns) in $T_\mathcal{P}$. Thus, the total number of
branches is $l = n + m$. In most cases, the number of forward branch instances exceeds the number of
backward ones in $T_\mathcal{P}$ (i.e., $n > m$), and there are significantly more forward branch
instances in $T_\mathcal{P}$ than there are basic blocks in $G_\mathcal{P}$ (i.e., $n \gg |V|$).

\begin{table*}
	\centering
	\small
	\resizebox{\textwidth}{!}{
	\begin{threeparttable}
		\begin{tabular}{c||c|c|c||c||c|c|c}
			\hline
			& \multicolumn{3}{c||}{Complexity Analysis} & \multirow{2}{*}{\shortstack{Context \\ Switch Triggers}} & \multicolumn{3}{c}{Evaluation Setup} \\ \cline{1-4} \cline{6-8}
			 & Trace$^\dagger$ & Measurement$^\dagger$ & Verification$^\ddagger$ & & CPU & Apps (optimization levels) & Evaluation Size  \\ \hline		
			Naive trace-based & $O(l)$& $O(1)$ & $O(l)$ &Forward \& Backward  & n/a & n/a & n/a \\ \hline
			OAT~\cite{sun2020oat} & $O(n)$ & $O(1)$ & $O(l|E|)$ &Forward \& Backward & Multi-core Cortex-A & 5 IoT apps including syringe pump~\cite{syringe} (\texttt{O0}) & $n < 1,000$ \\ \hline		
			C-FLAT~\cite{abera2016c} & n/a & $O(2^{|V|})$ & $O(2^{|V|})$ & Forward \& Backward  & Multi-core Cortex-A & Syringe pump (\texttt{O0}) & $|E|=322$ \\ \hline 
			LO-FAT~\cite{dessouky2017fat} & n/a & $O(2^{|V|})$ & $O(2^{|V|})$ & n/a & Customized multi-core RISC-V & Syringe pump (\texttt{O0}) & $|E|=322$ \\ \hline 			
			Blast~\cite{yadav2023whole} & $O(2^{|V|})$ & $O(1)$ & $O(2^{|V|})$ & Forward \& Backward & Multi-core Cortex-A & Syringe pump and Embench~\cite{Embench} (\texttt{O0}) & $|V|<987$\\ \hline	\hline			
			\enola & $O(|V|)$ & $O(1)$ & $O(2^{|V|})$ & Forward & Single-core Cortex-M & Syringe pump, Embench, and wolfSSL~\cite{wolfSSL}(\texttt{O0}, \texttt{O2}, \texttt{O3}, and \texttt{Oz}) & $|V|<5,480$\\ \hline
		\end{tabular}
	\end{threeparttable}
	}
	\caption{Space$^\dagger$/time$^\ddagger$ complexity, context switch triggers, and evaluation setup in C-FLAT, LO-FAT, OAT, Blast, and \enola.} 
	\label{t:criteria}
	\vspace{-0.3cm}
\end{table*}

\textbf{Full-trace based approach}.  In a naive trace-based attestation scheme, the authenticator
$Auth$ consists of the full trace $T_\mathcal{P}$, and the measurement is simply the cryptographic
hash of the full trace. The combined space complexity of the trace and the measurement is $O(l)$.
Subsequently, the verifier $\mathbb{V}$ performs an abstract execution of $\mathcal{P}$. During
abstract execution, whenever a non-sequential control-flow transfer site is encountered,
$\mathbb{V}$ verifies whether the next record in $T_\mathcal{P}$ belongs to the destination set of
the site $i$, denoted as $\cup e_i.d$.  The time complexity for verification is $O(l)$ as well.
However, due to the theoretically unbounded nature of $l$ (e.g., in the presence of an infinite
loop) and its practically substantial size, this approach is impractical, even for small programs
running on resource-constrained embedded systems.

\textbf{OAT}. To reduce the trace size, OAT~\cite{sun2020oat} employs three techniques: (1) a single bit to represent for each conditional branch outcome.
Compared to recording the entire address, this approach reduces the size while maintaining the same asymptotic space complexity; (2) for forward indirect branches (i.e., indirect jumps and indirect calls), OAT records the destination address; and (3) for backward edges maintains a single chained hash for return addresses, denoted as $H = hash(H \oplus RetAddr)$.
For the hash function, OAT uses a software implementation of BLAKE2s~\cite{BLAKE2}.
During abstract execution, $\mathbb{V}$ tracks the branch direction using the bit to determine the conditional branch path, validates destination addresses for indirect branches, and calculates and verifies hashes for return
instructions.
OAT reduces the trace and measurement size complexity to $O(n)$, while the
verification time complexity for $\mathbb{V}$ in OAT is $O(l|E|)$.

\textbf{C-FLAT and LO-FAT}. In C-FLAT~\cite{abera2016c} and LO-FAT~\cite{dessouky2017fat}, the $Auth$ consists solely of measurements, excluding any trace information entirely
(i.e., $T = \varnothing$). An exemplary $Auth$ takes the form of $(H_1, {\langle H_2, 5 \rangle,
\langle H_3, 4 \rangle, ...}, ...)$, where $H_1$ represents the cumulative hash of a path without
loops, $\langle H_2, 5 \rangle$ signifies the cumulative hash of a path inside a loop executed five
times, and $\langle H_3, 4 \rangle$ represents the cumulative hash of a different path inside the
same loop executed four times.
With this approach, if $\mathcal{P}$ has no loops, the measurement consists of only a single value,
resulting in a space complexity of $O(1)$. However, when loops are present, the measurement space
complexity in C-FLAT and LO-FAT grows to $O(2^{|V|})$, as it is bounded by the number of possible
paths within loops, which is itself bounded by $O(2^{|V|})$.
Additionally, because no trace information is available, the verifier $\mathbb{V}$ must explore all
possible paths to verify $Auth$, leading to a time complexity of $O(2^{|V|})$. For the hash
function, C-FLAT employs a software implementation of BLAKE2, while LO-FAT
incorporates a hardware SHA-3 engine directly into a RISC-V MCU.

\textbf{Blast}. To further reduce the trace size, Blast~\cite{yadav2023whole} does not generate
traces at the basic block level; instead, it operates at the function level. Each entry in the trace
takes the form $(Func, Path)$, where $Func$ represents the name of a function, and
$Path$ denotes the acyclic Ball-Larus path number~\cite{ball1996efficient} within that function.
To handle loops, Blast employs Ball-Larus' technique of resetting the path number to a non-zero
value at the back-edge of the loop. This approach effectively divides the Control Flow Graph (CFG)
into a series of acyclic components, with each component independently computing path numbers.
Consequently, for a CFG with loops, the size of a function’s trace is represented as a set of path
numbers, which in the worst case corresponds to the number of paths in the function, bounded by
$O(2^{|V|})$.
The measurement is computed as a hash of the trace.
Blast utilizes a software implementation of BLAKE2s as the hash function.

\textbf{Limitations of existing schemes}. As shown in Table~\ref{t:criteria}, due to the substantial
size of the trace or measurement data—e.g., $O(n)$ trace in OAT, $O(2^{|V|})$ measurement in C-FLAT and LO-FAT, and the $O(2^{|V|})$ trace in Blast, existing approaches can only handle small programs or even snippets. 
Note that our complexity analysis provides an upper bound.
In practice, real-world programs typically do not exhibit the worst-case complexity. Nonetheless, the space complexity of authenticators remains a fundamental limitation, affecting the scalability of prior approaches.
For instance, OAT was evaluated on operations containing fewer than 1,000 forward branches ($n$),
C-FLAT was tested on programs with up to 322 edges ($|E|$), and Blast was evaluated on programs with
at most 987 basic blocks ($|V|$).
Moreover, LO-FAT requires modifications to the CPU, making it incompatible with off-the-shelf devices.

%% file: background.tex
\section{Hardware Primitives on MCU}
In this section, we discuss the ARMv8-M MCU architecture, on which we implemented and evaluated \enola.
Note that \enola's design principles generalize to other architectures with comparable security extensions.
Furthermore, its linear-space transmission property for the attested program is architecture-agnostic.

\textbf{ARMv8-M Architecture and TrustZone}. The ARMv8-M architecture provides a 32-bit physical address space and features 16 general-purpose registers, namely \texttt{r0} to \texttt{r15}.
Among these, \texttt{r13}/\texttt{sp} serves as the stack pointer,
\texttt{r14}/\texttt{lr} (link register) holds the return address during subroutine calls, and
\texttt{r15}/\texttt{pc} is the program counter. 
ARMv8-M includes a trusted execution environment called TrustZone.
The TrustZone divides the memory into secure and non-secure states, where a memory region can be designated as secure, non-secure callable (NSC), or non-secure.
Secure state components can directly access non-secure resources, while the NSC region acts as a bridge for transitioning from the non-secure to secure state, facilitated by the \texttt{sg} (Secure Gateway) instruction.

\textbf{Memory Protection Unit}.
The Memory Protection Unit (MPU) in ARMv8-M architecture enables privileged software to partition memory regions and assign region-specific access and execution permissions.
For example, it can enforce the (W$\oplus$X) property by marking writable data as non-executable and code regions as read-only.
Any violation of these permissions triggers a memory management fault.

\textbf{ARMv8.1-M Pointer Authentication}. 
The Pointer Authentication (PA) extension includes several instructions, such as
\texttt{pacg}.
These instructions generate a keyed Pointer Authentication Code (PAC) for a pointer or data using the QARMA block
cipher~\cite{avanzi2017qarma}. The MCU provides four key registers for different use
cases.
For example, when the \texttt{pacg} instruction is executed in the unprivileged level and non-secure state, the
\texttt{pac\_key\_u\_ns} register is implicitly used as the key to compute the PAC. These key registers
can only be modified using the privileged \texttt{msr} (move-to-system-register) instruction.
Additionally, the secure state privileged code can modify both keys associated with the non-secure
state.

%% file: approach.tex
\section{\enola}

In this section, we first outline the system and threat model, followed by the functionality and workflow of \enola.

\input{threatmodel}

\subsection{\enola Trace and Measurement Schemes}
\label{s:trace}

In \enola, the authenticator $Auth = (T, M)$ includes a basic block occurrence trace ($T_{\mathcal{O}}$) and two measurements $M = \langle M_f, M_b \rangle$, representing the forward path ($M_f$) and backward path ($M_b$) taken by $\mathbb{P}$. Here, the backward path specifically refers to the sequence of function returns.
$T_{\mathcal{O}}$ is not a traditional trace of sequential operations; instead, it is structured as follows:
{
\setlength{\abovedisplayskip}{5pt} 
\setlength{\belowdisplayskip}{5pt} 
\begin{displaymath}
\scalebox{.83}{$T_{\mathcal{O}} = (\{\langle v_i.s, \#v_i \rangle | i = 0,..., |V|-1 \wedge \#v_i \neq 0\}, \{t_i| t_i \notin \bigcup_{i=0}^{|V|-1} v_i.s \})$}
\end{displaymath}
}
Here, $v_i.s$ represents the start address of the basic block $v_i$, and $\#v_i$ denotes the
occurrence count of $v_i$ during an execution. When $\#v_i$ equals zero, the pair $\langle v_i.s,
\#v_i \rangle$ is omitted from $T_{\mathcal{O}}$. Legitimate indirect branch or call targets
correspond to the start addresses of all basic blocks. However, in cases involving non-legitimate
targets, such as the middle of an instruction in ROP attacks, $T_{\mathcal{O}}$ can include a list
of these target addresses in ${t_i \mid t_i \notin \bigcup_{i=0}^{|V|-1} v_i.s }$.  Essentially, any
appearance of $t_i$ in $T_{\mathcal{O}}$ represents a control-flow violation. In practice, $|{t_i
\mid t_i \notin \bigcup_{i=0}^{|V|-1} v_i.s }|$ is a small number, and \enola sets a maximum threshold to ensure the constant size of this component.
Therefore, the trace space complexity for $\mathbb{P}$ is linear with respect to the number of basic blocks, i.e., $O(|V|)$.

The measurement $M$ is composed of two cumulative measurements for the forward path, i.e., $M_f$ and the backward path, i.e., $M_b$, respectively.
We calculate both measurements using the same hash chain approach:
$H_{K_m}$ represents the measurement function with the key of $K_m$.
For $M_f$, the destination address is denoted by $d_i$, while for $M_b$, $d_i$ represents the return address. 
{
\setlength{\abovedisplayskip}{5pt} 
\setlength{\belowdisplayskip}{0pt} 
\begin{displaymath}
	\scalebox{.9}{$M_i = \begin{cases}
		H_{K_m}(0, d_i) &\text{if $i = 0$}\\
		H_{K_m}(M_{i-1}, d_i) &\text{if $i > 0$}\\
	\end{cases}$}
\end{displaymath}
}

\begin{figure}[t]
	\includegraphics[width =\columnwidth]{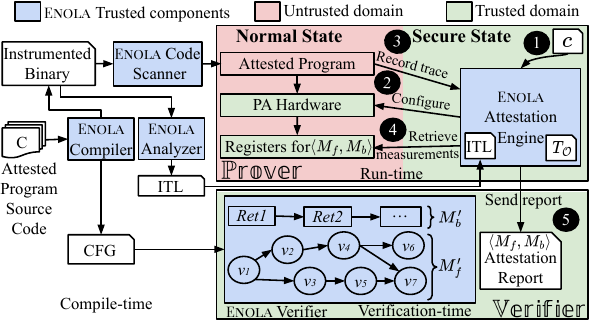}
	\caption{The workflow of \enola} 
	\label{fig:overview}
	\vspace{-0.6cm}
\end{figure}

\noindent\textbf{Attestation granularity.}
The authenticator design of \enola is independent of the selected instrumentation granularity.
In this work, we employ basic-block-level occurrence reporting and measurement generation to provide fine-grained control-flow attestation. Alternatively, the same representation may be instantiated at a coarser granularity, such as function-level attestation similar to Blast~\cite{yadav2023whole}, to trade verification granularity for lower instrumentation and runtime overhead while retaining bounded transmitted-authenticator size with respect to the selected attestation units. Evaluating such granularity variants is outside the scope of this work.
\vspace{-0.2cm}
\subsection{\enola Components and Workflow}
\label{s:overview}

The workflow (shown in Figure~\ref{fig:overview}) of \enola consists of three distinct stages: compile-time, run-time, and verification-time.

\textbf{Compile-time}.
\emph{\enola Compiler}: The \enola compiler instrumentation serves two primary purposes: (1) Reporting Control-Flow Events: The instrumentation reports control-flow transfer events in the program to the attestation engine, which operates in a secure state. This enables the attestation engine to construct $T_{\mathcal{O}}$, a trace of the program's execution. The details of $T_{\mathcal{O}}$ construction are discussed in~\S\ref{subsec:tracegen}. (2) Calculating and Storing Measurements: The PA instruction instrumentations compute measurements and store them in designated general-purpose registers.
These measurements are further elaborated in~\S\ref{subsec:pacformeasure}.
Since the measurements are never stored in memory, they are inherently protected from memory corruption attacks. This approach ensures secure and reliable operation of the attested program.

Furthermore, the \enola compiler generates an overapproximated control-flow graph (CFG) that is utilized by the \enola verifier for abstract execution.
While generating a comprehensive CFG encompassing all possible control-flow transfers remains an open research challenge~\cite{carlini2015control, conti2015losing}, microcontroller programs ($\mathcal{P}$) are typically less complex than desktop applications.
This simplicity, combined with advancements in control-flow analysis techniques~\cite{lu2019does, kim2021refining}, supports the widely accepted assumption that the CFG for such systems can be treated as complete~\cite{abera2016c, yoo2022kernel, lu2023practical, bai2021static, yang2021demons, zhang2021recfa, cloosters2022riscyrop}.
This assumption forms the foundation for leveraging the CFG in the attestation and verification processes.

\emph{\enola Code Scanner}:
Since the compiled program may execute at the privilege level of the normal world, it presents a potential vulnerability where attackers could manipulate PA key registers or reserved general-purpose registers used for measurements. To mitigate this risk, \enola incorporates a code scanner to ensure the integrity of the execution environment.
The \enola code scanner performs a comprehensive analysis to verify that no programs operating in the normal state -- including the attested program, libraries, and the kernel -- contain any instructions or gadgets capable of tampering with these critical registers. This proactive approach ensures the security of the measurement process by preventing unauthorized modifications at the software level.

Specifically: (1) The code scanner ensures no \texttt{msr} instructions write to the PA key registers. (2) It detects and flags \texttt{mov} or \texttt{pop} instructions that could modify the two designated general-purpose registers used for measurements.
The scanner also flags handwritten assembly that contain assembly-level instructions that directly modify the reserved measurement registers or  PA key registers. 
If the scanner identifies such instructions, it issues a warning to the developer, indicating that their presence could compromise the security guarantees of \enola. This allows the developer to revise the source code and eliminate the problematic instructions. 
This methodology has been validated and adopted as a best practice in previous research studies~\cite{zhou2020silhouette,du2022kage}, further reinforcing its effectiveness in maintaining a secure execution environment.


\emph{\enola Analyzer}: 
The \enola analyzer facilitates efficient attestation by enabling the \enola attestation engine to determine whether the destination of an indirect branch corresponds to a valid basic block. This determination is crucial for deciding whether to update $\{\langle v_i.s, \#v_i \rangle\}$ or $\{t_i\}$ during execution.
To achieve this, the \enola analyzer performs both static and dynamic analysis on the instrumented binary to construct an Indirect Target List (ITL), which contains the valid destination addresses for all potential indirect branches or calls. 
The presence of the ITL eliminates the need for the attestation engine to dereference and validate the destination address of an indirect branch at run-time to determine whether it corresponds to the starting address of a legitimate basic block. This optimization significantly improves the performance and reliability of the attestation process.


\textbf{Run-time}.
\emph{\enola Attestation Engine}:
At run-time, the attested program $\mathcal{P}$ executes in the normal state, while the \enola attestation engine runs in the secure state.
At system boot, the attestation engine executes first, and receives a nonce $c$ from the remote verifier (\Circled{\textbf{1}}).
It then retrieves keys from secure storage and configures the PA key registers (\Circled{\textbf{2}}). 
During the configuration step, the attestation engine also enables the PA hardware to operate in the normal state and configures the Memory Protection Unit (MPU) to enforce W$\oplus$X.
As previously mentioned, during the execution of $\mathcal{P}$, instrumented instructions trigger the attestation engine to record the occurrence trace $T_\mathcal{O}$ (\Circled{\textbf{3}}).
Additionally, these instrumented instructions invoke the PA hardware to generate measurements and store the cumulative values $\langle M_f, M_b \rangle$ in reserved registers.
On the completion of $\mathcal{P}$, the attestation engine retrieves these values from the reserved registers (\Circled{\textbf{4}}).
Subsequently, along with $c$, it generates a signature over $Auth = (T_\mathcal{O}, \langle M_f, M_b \rangle)$, constructing the attestation report. This report is then transmitted to $\mathbb{V}$ for verification (\Circled{\textbf{5}}).

\textbf{Verification-time}.
\emph{\enola Verifier}:
Upon receiving the report from $\mathbb{P}$, $\mathbb{V}$ 
authenticates the signature using $c$ and $K_a$. The \enola verifier then abstractly executes $\mathcal{P}$, guided by the trace $T_\mathcal{O}$, while comparing the recalculated measurements with the received values.
The core component of $\mathbb{V}$ is a backtracking algorithm, which is discussed in detail in~\S\ref{s:verfi_algo}.


\definecolor{highlight1}{RGB}{204,255,255}
\colorlet{FancyVerbHighlightColor}{highlight1}

\begin{listing}[h]
	{\renewcommand\fcolorbox[4][]{\textcolor{gray}{\strut#4}}
		\begin{minted}[xleftmargin=5.8pt,numbersep=1pt, tabsize=2, frame=lines, framesep=1mm, breaklines, highlightlines={5-10,14-19}, linenos=true, escapeinside=||, fontsize=\scriptsize]{tasm}
	comparator:
		cmp	r0, #0x2
		bne	<branch_2>
	branch_1:
		|\xglobal\colorlet{FancyVerbHighlightColor}{green!10}|push {r0-r3, lr}
		|\xglobal\colorlet{FancyVerbHighlightColor}{highlight1}|mov	r0, pc
		add.w	r0, r0, #0xb
		pacg	r10, r0, r10 ;update measurement |$M_f$|
		|\xglobal\colorlet{FancyVerbHighlightColor}{green!10}|bl <report_direct> ;report occurrence trace
		pop {r0-r3, lr}
		ldr	r0, [sp, #0x4];branch-1 instructions|\xglobal\colorlet{FancyVerbHighlightColor}{highlight1}|
		...
	branch_2:
		|\xglobal\colorlet{FancyVerbHighlightColor}{green!10}|push {r0-r3, lr}
		|\xglobal\colorlet{FancyVerbHighlightColor}{highlight1}|mov	r0, pc
		add.w	r0, r0, #0xb
		pacg	r10, r0, r10 ;update measurement |$M_f$|
		|\xglobal\colorlet{FancyVerbHighlightColor}{green!10}|bl <report_direct> ;report occurrence trace
		pop {r0-r3, lr}
		ldr	r0, [sp]			;branch-2 instructions
		\end{minted}
	} 
	\cprotect\caption{Instrumentation example: direct branches. Light green: trace reporting, light blue: measurement calculation}
	\label{code:cond}
	\vspace{-0.2cm}
\end{listing}

\colorlet{FancyVerbHighlightColor}{highlight1}

\subsection{Secure Generation of $T_{\mathcal{O}}$}
\label{subsec:tracegen}


The \enola secure generation of $T_{\mathcal{O}}$ involves three steps: the attestation engine provides runtime branch destination reporting interfaces, the \enola compiler instruments interface calls, and the attestation engine logs the resulting trace.

\textbf{Non-secure callable branch destination reporting interface}.
The \enola attestation engine introduces non-secure callable functions: \texttt{report\_direct} and \texttt{report\_indirect}, to allow the attested program to report its branch destinations.
The \texttt{report\_direct} function does not require any parameters, as the attestation engine can directly infer the destination address of branch sites. 
By contrast, the \texttt{report\_indirect} function requires the destination address to be provided as a parameter in \texttt{r0}.
Both are annotated with the ARM Clang compiler \texttt{\_\_attribute\_\_((cmse\_nonsecure\_entry))} directive of the Cortex-M Security Extensions (CMSE). 
This directive instructs ARM Clang to produce a trampoline within the non-secure callable memory region and position the actual function within the secure memory region.
Although both the trampoline and the actual function share the same name, 
for the direct branch reporting case, they operate within distinct symbol scopes.
To report direct branches, the \enola compiler instruments the attested program to initiate calls (using the \texttt{bl} instruction) to the \texttt{report\_direct} trampoline, ensuring that the destination is preserved in the \texttt{lr} register.
The trampoline contains only two instructions: a secure gate and a direct branch. Neither instruction modifies the \texttt{lr} register.
 \enola attestation engine extracts the direct branch destination ($v_i.s$) of the attested program from the \texttt{lr} register, leveraging a \texttt{mov} instruction.
For indirect branches, the destination is available in \texttt{r0} for the attestation engine.



\textbf{Instrumenting Reporting: Direct Branches and Loops}.
For direct branches (e.g., from \texttt{if-else} or \texttt{switch} statements), 
\enola inserts ``\texttt{bl <report\_direct>}" instructions. These calls are placed at the beginning of each branch target basic block, such as \texttt{branch\_1} and \texttt{branch\_2}, as shown in Listing~\ref{code:cond}.
To reduce instrumentation overhead, \enola avoids instrumenting basic blocks that end with branch instructions, such as the \texttt{comparator} block and the \texttt{bne} instruction in Listing~\ref{code:cond}. Since these blocks immediately dominate their branch target basic blocks, additional tracing is unnecessary.
Listing~\ref{code:cond} illustrates an example where instrumented calls are inserted on lines 9 and 18 to report the taken path. To ensure correctness and avoid interference, the \enola instrumentation surrounds these calls with instructions to store and restore the caller-saved registers, along with the \texttt{lr} register, to or from the stack. This precaution is particularly important because these registers may be used as general-purpose registers, especially under \texttt{O2} or \texttt{Oz} compiler optimizations (as shown in lines 5, 10, 14, and 19).

For loops, \enola instruments the loop body and exit basic blocks as shown in the example in Listing~\ref{code:loop}.
The loop condition block is not instrumented because it is the immediate dominator of both the body and exit blocks.
This method inherently accounts for the \texttt{break} statement, as the block it is in is immediately post-dominated by the exit block.



\begin{listing}[h]
	{\renewcommand\fcolorbox[4][]{\textcolor{gray}{\strut#4}}
		\begin{minted}[xleftmargin=5.8pt,numbersep=1pt, tabsize=2, frame=lines, framesep=1mm, breaklines, highlightlines={7-12,17-22}, linenos=true, escapeinside=||, fontsize=\scriptsize]{tasm}
		ldr	r0, [sp, #0x4]	;loop counter
		ldr	r1, [sp, #0x10] ;loop limit
	loop_condition:
		cmp	r0, r1
		bge	<loop_exit>
	loop_body:
		|\xglobal\colorlet{FancyVerbHighlightColor}{green!10}|push {r0-r3, lr}
		|\xglobal\colorlet{FancyVerbHighlightColor}{highlight1}|mov	r0, pc
		add.w	r0, r0, #0xb
		pacg	r10, r0, r10 ;update measurement |$M_f$|
	|\xglobal\colorlet{FancyVerbHighlightColor}{green!10}|	bl <report_direct> ;report occurrence trace
		pop {r0-r3, lr}
		...		 				 	;loop-body instructions
		adds	r0, #0x1 |\xglobal\colorlet{FancyVerbHighlightColor}{highlight1}|
		b	<loop_condition>
	loop_exit:
		|\xglobal\colorlet{FancyVerbHighlightColor}{green!10}|push {r0-r3, lr}
		|\xglobal\colorlet{FancyVerbHighlightColor}{highlight1}|mov	r0, pc
		add.w	r0, r0, #0xb
		pacg	r10, r0, r10 ;update measurement |$M_f$|
	|\xglobal\colorlet{FancyVerbHighlightColor}{green!10}|	bl <report_direct> ;report occurrence trace
		pop {r0-r3, lr}
		add	sp, #0x18		 ;loop-exit instructions
		\end{minted}
	} 
	\cprotect\caption{Instrumentation example: loops}
	\vspace{-0.2cm}
	\label{code:loop}
\end{listing}

\colorlet{FancyVerbHighlightColor}{highlight1}
\vspace{-0.2cm}
\textbf{Instrumenting Reporting: Indirect Branches}.
Unlike direct branches, which have statically determined targets, indirect branches can be manipulated to jump to arbitrary locations, including potentially the middle of instructions.
As a result, in addition to instrumenting the start of target basic blocks, as is done for direct branches, the \enola compiler must also instrument indirect calls or jumps to capture the destination address. The destination addresses of indirect branches cannot be derived from the \texttt{lr} register. Instead, they may involve any general-purpose register used by the \texttt{blx} or \texttt{bx} instructions.
Listing~\ref{code:indirect} provides an example of instrumenting an indirect call site (Line 8), where the trampoline call receives the target address via the \texttt{r0} parameter. 
The process involves saving caller-saved registers (\texttt{r0} - \texttt{r3}) to the stack, then transferring the target address into \texttt{r0} (Line 4) to set up for the \texttt{report\_indirect} trampoline invocation (Line 6). 
The instrumented sequence concludes with restoring the caller-saved registers from the stack (Line 8).
\begin{listing}[h]
	{\renewcommand\fcolorbox[4][]{\textcolor{gray}{\strut#4}}
		\begin{minted}[xleftmargin=5.8pt,numbersep=1pt, tabsize=2, frame=lines, framesep=1mm, breaklines, highlightlines={3-7}, linenos=true, escapeinside=||, fontsize=\scriptsize]{tasm}
	movw	r3, r8
	movt	r3, r9
	|\xglobal\colorlet{FancyVerbHighlightColor}{green!10}|push	{r0-r3}			 ;store caller-saved registers
	mov	r0, r3				 ;copy destination to r0 
	|\xglobal\colorlet{FancyVerbHighlightColor}{highlight1}|pacg	r10, r0, r10	;update measurement |$M_f$|
	|\xglobal\colorlet{FancyVerbHighlightColor}{green!10}|bl <report_indirect>;report occurrence trace 
	pop	{r0-r3}				;restore caller-saved registers
	blx	r3		 				;indirect call site
		\end{minted}
	} 
	\cprotect\caption{Instrumentation example: indirect calls}
	\vspace{-0.1cm}
	\label{code:indirect}
\end{listing}

\colorlet{FancyVerbHighlightColor}{highlight1}
\vspace{-0.7cm}
\subsection{Secure and Efficient Measurement Calculation}
\label{subsec:pacformeasure}




\textbf{Measurement key initialization}.
The \enola attestation engine retrieves the measurement key $K_m$ from the secure storage and loads it into the PA key registers \texttt{pac\_key\_u\_ns} and \texttt{pac\_key\_p\_ns} with the privileged \texttt{msr} instruction.
This component should be implemented in assembly, leveraging general-purpose registers to temporarily transfer the keys from secure storage to the PA key registers. As a result, the measurement key $K_m$ is never spilled to memory.

\textbf{Reserving general-purpose registers}.
To prevent measurements from being spilled to memory, the \enola compiler reserves the general-purpose registers \texttt{r10} and \texttt{r11} to securely store $M_f$ and $M_b$, respectively. Both registers are initialized to zero before the attestation engine transfers control to the non-secure state.
These registers are specifically chosen because they are the callee-saved registers with the highest numerical identifiers in the ARM Procedure Call Standard. Consequently, apart from the instrumented measurement instructions, the compiled REE program-including the attested program-does not utilize these registers, ensuring that the cumulative measurements are not spilled to memory.

\textbf{Robustness of register reservation}.
The \enola compiler removes \texttt{r10} and \texttt{r11} from the register allocation pool, making the reservation persistent across compiler optimizations or other flags, including aggressive optimization settings.
The implementation details of this compiler  reservation are discussed in~\ref{s:impl}.
Furthermore, as \texttt{r10} and \texttt{r11} are callee-saved registers, the REE program produced by this method is compatible with programs not compiled by the \enola compiler, including pre-compiled or vendor libraries.
If any uninstrumented program uses these registers, the calling convention will restore their original values upon return.
Thus those uninstrumented programs cannot directly influence the measurements.
However, such usage could cause the measurements to be spilled into memory, exposing them to potential memory corruption attacks.

\textbf{Instrumenting measurement calculation: forward path}.
\enola utilizes the \texttt{pacg} instruction to compute the measurements. Similar to the instrumentation used for direct branch reporting, the \enola compiler inserts PA instructions at the start of all destination basic blocks for forward branches. This is illustrated by the light blue instructions in Listings~\ref{code:cond} and~\ref{code:loop}.
Specifically, the instrumentation retrieves the program counter value into a free general-purpose register (e.g., \texttt{r4}) and subsequently increments it by a predetermined value to obtain the target basic block address ($v_i.s$) (Lines 6-7 in Listing~\ref{code:cond}).
Then, a \texttt{pacg} instruction is instrumented to sign the value in the available general-purpose register, using the previous measurement in \texttt{r10} as the modifier (Line 8).
Similar to the instrumentation for indirect branch reporting, 
the \enola compiler instruments the indirect branch sites with PA instructions.
Listing~\ref{code:indirect} shows an example \texttt{pacg} instrumentation (Line 5) for an indirect call site, where the destination is already copied to the \texttt{r0} register for trace reporting. 
\vspace{-0.2cm}
\begin{listing}[h]
	{\renewcommand\fcolorbox[4][]{\textcolor{gray}{\strut#4}}
		\begin{minted}[xleftmargin=5.8pt,numbersep=1pt, tabsize=2, frame=lines, framesep=1mm, breaklines, highlightlines={7-8}, linenos=true, escapeinside=||, fontsize=\scriptsize]{tasm}
	prologue:
		push	{r7, lr}
		sub	sp, #0x28
		...
	epilogue:
		add	sp, #0x28
		ldr	r4, [sp, #0x4]
		pacg	r11, r4, r11	 ;update measurement |$M_b$|
		pop	{r7, pc}
		\end{minted}
	} 
	\vspace{-0.2cm}	
	\cprotect\caption{Instrumentation example: non-leaf return} 
	\label{code:nlret}
\end{listing}

\vspace{-0.3cm}
\textbf{Instrumenting measurement calculation: backward path}.
The \enola compiler instruments \texttt{pacg} instruction before all function returns to construct the $M_b$ measurements.
In the Cortex-M architecture, non-leaf functions preserve return addresses on the stack by pushing \texttt{lr}, as shown in Line 2 of Listing~\ref{code:nlret}. Subsequently, function returns are executed by directly popping the saved return address into \texttt{pc}.
For non-leaf functions, \enola instrumentation first loads the return address from the stack into a free general-purpose register. It then inserts a \texttt{pacg} instruction to compute $M_b$, storing the result in the \texttt{r11} register (Lines 7 and 8).

Conversely, leaf functions retain the return address in \texttt{lr} without spilling it to the stack.
They return via the "\texttt{bx lr}" or "\texttt{mov pc, lr}" instructions. 
Leaf function  returns are directly instrumented using \texttt{"pacg r11, lr, r11"}.
Importantly, \enola does not report the occurrence trace for the backward path, thereby \textbf{eliminating the need for a context switch} to the attestation engine in the secure state.



\subsection{Secure ISR Dispatcher}
Given the frequent and unpredictable nature of interrupts in embedded systems and the fact that they do not conform to a predetermined CFG, \enola employs a runtime enforcement, a modified version of ISC-FLAT~\cite{neto2023isc}, to ensure ISRs always return benignly to the program $\mathcal{P}$.
\enola incorporates a secure state interrupt dispatcher that intercepts all normal state interrupts.
Leveraging the TrustZone-protected Nested Vectored Interrupt Controller (NVIC), it interposes itself between the interrupt trigger and ISR execution.
Upon interception, it saves the context of $\mathcal{P}$: including the stack pointer, measurements $M_f$ and $M_b$, and normal state system registers onto the secure stack.
Before transferring control to the untrusted ISR, the dispatcher configures the normal state MPU to mark $\mathcal{P}$’s stack region as read-only, thereby preventing any unauthorized writes by the ISR.
Crucially, this allows the system to maintain control-flow integrity without including the ISRs in the TCB.
On ISR completion, the dispatcher restores the saved context to ensure the correct resumption of $\mathcal{P}$.
It also reconfigures the MPU to mark $\mathcal{P}$'s stack region as both readable and writable.

On interrupt entry, \enola dispatcher records a flag $I_x$ in the occurrence trace $T_{\mathcal{O}}$ and it's cleared upon ISR's return to the dispatcher.
Thus, presence of $I_x$ in $T_{\mathcal{O}}$ signals a deviation from expected ISR behavior, i.e., indicating malicious execution in the normal state interrupt handler or the program $\mathcal{P}$.
Notably, $I_x$ preserves $Auth$'s linear space complexity, as it maintains constant space in $T_{\mathcal{O}}$  when a compromised ISR executes.
\vspace{-0.2cm}
\subsection{Backtracking Algorithm for Verification}
\label{s:verfi_algo}
The \enola verifier employs the backtracking algorithm to verify the legitimacy of the attested control path. The algorithm takes as inputs the attestation report $\mathcal{R}$ and the control-flow graph $G_\mathcal{P}$ of the attested program, which includes the entry basic block ($\mathcal{P}_{entry}$) and the potential exit points ($\mathcal{P}_{exits}$).


The algorithm begins by verifying the attestation report signature and checking for any illegal indirect branch targets.
Upon successful verification, it abstractly executes the program using the $G_\mathcal{P}$ and validates both the forward and backward measurements. The execution starts at $\mathcal{P}_{entry}$, initializes an empty simulated call stack to track function returns, and recursively executes branches as guided by $Auth$.

Based on the last instruction ($v_{c}.e$) within the current basic block ($v_{c}$), the algorithm executes one of the following:
(1) Program Exit: For exits at $v_{c}.e$, the verifier concludes execution and compares the computed measurements with the received values in $Auth$.
(2) Function Call: For a function call at $v_{c}.e$, the verifier pushes the next instruction address onto the simulated call stack and continues execution at the call target.
(3) Conditional Branch: If $v_{c}.e$ is a conditional branch, the algorithm explores all potential paths emanating from the current basic block after verifying against $T_\mathcal{O}$. A non-zero value in the occurrence count signifies a valid transition to that target. The algorithm decrements count, updates the forward measurement, and proceeds along the path. If no valid targets exist in $T_\mathcal{O}$, the path is deemed invalid, and the algorithm backtracks.
(4) Return Instruction: For returns, the backward measurement is updated, and execution continues at the return target, obtained from the simulated function call stack.

\textbf{Complexity Analysis.} The verificaiton algorithm has an exponential worst-case verification bound, $(O(2^{|V|}))$, because the verifier may need to backtrack when multiple paths are consistent with the same basic-block occurrence counts.
The largest benchmark, wolfSSL/RSA, has 5,480 basic blocks. 
In practice, this bound is overly conservative: the verifier usually follows only a few count-compatible successors, while ($M_f$), ($M_b$), and the simulated call stack prune inconsistent paths.
Hence, practical verification is better viewed as ($O(2^k \cdot L)$), where (L) is the realized abstract execution length and (k) is the number of unresolved regions where multiple paths are consistent with $T_\mathcal{O}$.
Significant backtracking is possible in synthetic or highly symmetric CFGs, but it does not dominate realistic embedded workloads such as wolfSSL/RSA.

%% file: threatmodel.tex
\subsection{System and Threat Model}
\label{s:threat}
\textbf{System model}. \enola operates under the assumption that the embedded system processor provides a TEE, an MPU, and hardware capabilities for keyed message authentication code computations within the REE.
The attested program can execute at unprivileged or privileged level within the REE, while the \enola
attestation engine operates within the TEE.
The verifier in \enola can reside on any system or device, such as x86, Cortex-A, etc., but is assumed to be on a powerful machine or cloud-based system to ensure fast verification.

\enola's hardware assumptions are practical for modern security-oriented MCUs.
TrustZone was introduced to the M-profile through the ARMv8-M Security Extension and is available in Cortex-M23, Cortex-M33, and subsequent security-enabled implementations~\cite{ARMTz1}.
MPU support predates ARMv8-M and is available in Cortex-M3 and later Cortex-M implementations~\cite{ARMTz1}.
Key MAC was introduced first in the A-profile with ARMv8.3-A and later extended to the M-profile through the ARMv8.1-M PACBTI extension, implemented in recent processors such as Cortex-M85~\cite{ARMTz1}.
In \enola, the secure attestation engine configures and manages TrustZone, MPU permissions, and PA keys at runtime; this engine is developed as part of the framework and constitutes its TCB.

\textbf{Threat model}. We assume the presence of secure boot to ensure both: (1) the \enola attestation engine's code and data, and (2) the REE software, are securely loaded at boot time.
Not every piece of code within the REE requires attestation; the portion that undergoes attestation is referred to as the attested program.
We implement code immutability, e.g., W$\oplus$X, for the attested program.
While the \enola attestation engine is trusted during runtime, the control flow of the REE software, including the attested program, could be compromised at runtime.
Interrupt Service Routines (ISRs) in the REE are also outside the Trusted Computing Base (TCB) and may be compromised, containing control flow hijacking vulnerabilities (e.g., ROP gadgets).
Although REE ISRs are untrusted, \enola detects malicious ISR executions by monitoring ISR transitions and reflecting ISR-induced deviations in the attestation report.
The prover and verifier share both the measurement key ($K_m$) and attestation key ($K_a$), which can be practically established through manufacturing-time provisioning, device enrollment, or secure bootstrapping.
The measurement key is used for calculating measurements, while the attestation key is employed to sign the attestation report.
We assume secure storage is available to protect both the measurement key and attestation key at rest.
However, attackers could attempt to compromise memory content within the REE, use ROP gadgets to manipulate general-purpose registers, and modify key registers.
TOCTOU attacks~\cite{de2021toctou} and physical attacks such as power analysis or electromagnetic analysis, timing attacks, and data-only attacks considered out of scope for \enola.

%% file: analysis.tex
\section{Security Analysis}

To successfully hijack the control flow and bypass \enola's monitoring, an attacker must meet six key attack prerequisites:
(P1): Disable or bypass the instrumented code.
(P2): Tamper with \enola measurements $M_f$ and $M_b$, or the measurement key $K_m$.
(P3): Execute malicious control flow to generate hash collisions.
(P4): Exploit non-secure callable trampoline interfaces in the attestation engine.
(P5): Replay or corrupt $Auth$, $T_{\mathcal{O}}$, or the attestation key $K_a$ to manipulate verification at $\mathbb{V}$.
(P6): Exploit vulnerable ISRs in REE to tamper with $\mathcal{P}$'s stack, return address, or $M_f$ and $M_b$, thereby disrupting control-flow integrity or trace correctness.

P1 is mitigated through the enforcement of code immutability via the MPU~\cite{clements2017protecting,kim2018securing}; and any attempt to bypass the instrumentation or interface calls would be reflected in measurements $M_f$ and $M_b$.
For P2, the measurements $M_f$ and $M_b$ or PA key $K_m$ are never spilled into memory and stored in reserved registers accessible only by the instrumented \texttt{pacg} instructions.
Since the program source code is compiled with the \enola compiler, these measurement registers are excluded from the program binary, even under aggressive optimizations (e.g., \texttt{O3}, \texttt{Oz}).
The compiled binary is vetted by the \enola code scanner, which flags unauthorized instructions, handwritten assembly, or gadgets that could tamper with the reserved or PA key registers.

Although an attacker can exploit application vulnerabilities for control-flow violations, these are detected through trace and measurement verification.
Evading detection would require to forge hash collisions (P3), which is highly infeasible as QARMA block cipher has a collision probability of $2^{-60}$~\cite{avanzi2017qarma}.
\enola's design further makes valid collisions harder by chaining control-flow measurements as modifiers, given  \textit{N} transitions, resulting in a cumulative bound of $N*2^{-60}$.

\enola prevents direct manipulation of $Auth$, $T_{\mathcal{O}}$, or $K_a$ by storing and signing them in the secure state or PA register, while the random nonce $c$ thwarts replay attacks.
With $M_f$ and $M_b$ confined to REE registers, the failure to achieve P2, combined with the trusted ARM PA hardware for measurement computation, ensures the integrity of measurements in $Auth$.
Thus, exploiting trampoline interface calls becomes the only viable option to manipulate $Auth$ or $T_{\mathcal{O}}$.
However, \enola compiler prohibits any direct world-switch calls targeting the attestation engine interfaces, allowing such calls only at instrumented locations.
As a result, attackers are prevented from achieving P4, and when combined with TEE and ARM PA security guarantees, P5 is also effectively mitigated.

Normal state ISRs may contain memory corruption bugs exploitable for control flow hijacking (e.g., ROP) or tampering with attested program execution by invalid returns or manipulating measurements.
\enola mitigates this with a secure ISR dispatcher that intercepts all interrupts, saves context and measurements in secure memory, and marks $\mathcal{P}$’s stack as read-only during ISR execution to block unauthorized writes.
If the ISR fails to return to the dispatcher, a flag is recorded in $T_{\mathcal{O}}$, signaling malicious behavior and mitigating P6.

\section{Residual Risks}
\label{sec:residual-risks}

\enola's guarantees hold under the threat model in Section~\ref{s:threat}: secure boot for the attestation engine and REE software, correct TrustZone/MPU configuration, secure storage for $K_m$ and $K_a$, and an uncompromised attestation engine at runtime.
Compromise of the TrustZone TCB, secure-world software, MPU-enforced W$\oplus$X or $\mathcal{P}$'s stack , unsafe DMA/peripheral configuration, microarchitectural attacks, or TOCTOU attacks is outside scope of \enola.

\enola's correctness also depends on the quality of the recovered CFG.
An incomplete CFG may reject legitimate executions, while an over-approximated CFG may include infeasible paths.
Our implementation conservatively favors over-approximation to preserve MCU functionality while relying on CFG analysis to minimize infeasible paths.

\enola detects abnormal ISR behavior, including invalid returns and attempts to corrupt $\mathcal{P}$'s stack or measurement state, but does not reconstruct the exact malicious ISR path; such behavior is reflected in $T_O$.
\enola supports mixed binaries, including vendor libraries, through memory-based restoration of reserved measurement registers, improving deployability at the cost of potential memory-corruption risks and weaker register-based isolation.
These residual risks define \enola’s trust boundary: it provides control-flow attestation for the attested program under a securely booted and correctly configured TrustZone/MPU-based TCB, secure key storage, uncompromised measurement logic, and a sound CFG representation.



%% file: implementation.tex
\section{Implementation}
\label{s:impl}


We developed a prototype of \enola for the ARMv8.1-M architecture. The \enola LLVM compiler modules comprise 4,675 lines of C++ code.
It incorporates front- and back-end passes to annotate, instrument, and adjust code for accurate, optimization-resilient measurements.
The \enola attestation engine consists of 307 lines of C and inline assembly code.
The \enola analyzer consists of 138 lines of Python code. Additionally, the code scanner and verifier are implemented with 93 and 694 lines of Python code, respectively, leveraging the angr and pyelftools libraries.

\begin{figure}[h]	
	\includegraphics[width = .48\textwidth]{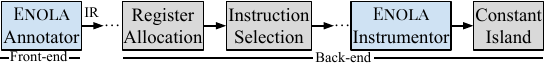}
	\caption{\enola passes in the LLVM pass pipeline} 
	\label{fig:llvm}	
	\vspace{-0.15cm}	
\end{figure}

The \enola compiler extends the LLVM 16.0.0 ARM embedded toolchain with the front-end \enola Annotator and back-end \enola Instrumentor passes.
We also modify \texttt{ARMBaseRegisterInfo} to reserve \texttt{r10} and \texttt{r11} for measurement by excluding them from LLVM's physical register allocator.
Consequently, LLVM cannot use these registers for variables, temporaries, or spill/reload operations, and machine-code-level enforcement prevents \texttt{O2}, \texttt{O3}, and \texttt{Oz} optimizations from overriding this constraint.

The \enola Annotator marks functions and valid direct-branch target basic blocks with LLVM IR metadata.
The \enola Instrumentor then uses these annotations to insert forward-path measurement instructions and \texttt{report\_direct} trampoline calls.
It also instruments indirect branches with \texttt{report\_indirect} calls and inserts PA instructions before function returns to compute backward-path measurements.

To support \texttt{O2}, \texttt{O3}, and \texttt{Oz} optimizations, all instrumentation is performed by the back-end \enola Instrumentor, avoiding LLVM IR phi-node constraints.
Because \texttt{lr} may be used as a general-purpose register, we modify \texttt{ARMFrameLowering} to save and restore it for annotated non-leaf functions, as shown in Listing~\ref{code:cond}.
The Instrumentor also integrates with the \texttt{ARMConstantIsland} pass to adjust instrumentation addresses when immediate offsets are limited, as illustrated in Figure~\ref{fig:llvm}.
Running after instruction selection and register allocation enables precise and efficient instrumentation.

The \enola analyzer utilizes angr to identify valid target addresses for instructions inducing indirect control-flow changes. Additionally, it employs dynamic training in a controlled environment with benign inputs to uncover control-flow transfers that may have been overlooked by angr.
The analyzer ultimately produces an Indirect Target List (ITL) containing all destination addresses for all indirect calls and branches.
The \enola code scanner disassembles the generated binary using angr and scans for privileged \texttt{msr} instructions or Return-Oriented Programming (ROP) gadgets that could overwrite the ARM PA key or measurement registers.
\vspace{-.2cm}

%% file: evaluation.tex
\section{Evaluation}

\begin{figure}[t]	
	\includegraphics[width = .48\textwidth]{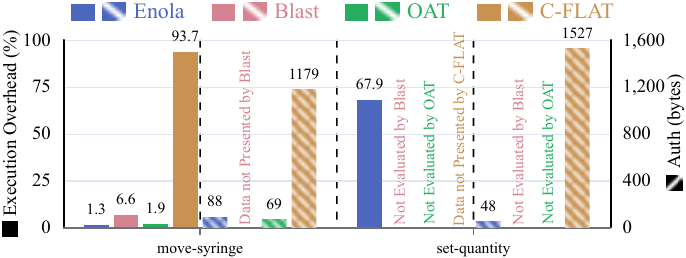}
	\vspace{-0.2cm}	
	\caption{Syringe pump: Execution time \& $Auth$ comparison}
	\label{fig:syring}
	\vspace{-.4cm}
\end{figure}

We present five key evaluations:
(1) a case study of a vulnerable embedded application demonstrating \enola’s security guarantees;
(2) an overhead analysis across 19 applications from Embench and wolfSSL;
(3) a comparative evaluation against existing solutions;
(4) an analysis of linearity in $Auth$ size; and
(5) a performance evaluation of the \enola secure ISR dispatcher.
Applications were evaluated with optimization \texttt{O0}, \texttt{O2}, \texttt{O3}, and \texttt{Oz}; we report results for \texttt{O2} and \texttt{Oz}.

\subsection{Evaluation Environment}

We evaluated \enola on the ARM Versatile Express Cortex-M prototyping FPGA system (V2M-MPS3)~\cite{ARMMPS3}. The system was configured as a Cortex-M85 microcontroller running at 25 MHz using the AN555 FPGA image (BSP v1.3.0).
\enola is the first solution (Table~\ref{t:criteria}) to be evaluated on single-core, low-end embedded CPUs, whereas previous solutions were all evaluated on multi-core, high-end, or mid-range CPUs.

\subsection{Evaluations on Syringe Pump Application}
We evaluated a version adapted from C-FLAT~\cite{abera2016c}, with minor source-code adjustments to support our compilation process for the Cortex-M85 microcontroller. The program has two main control-flow paths: the \texttt{move-syringe} path, which dispenses or withdraws the specified bolus amount (\texttt{+/-}); and the \texttt{set-quantity} path, which defines the bolus amount in milliliters from user input. Figure~\ref{fig:syring} compares the execution overhead and $Auth$ sizes for both control-flow paths.

\textbf{Execution time overhead and $Auth$ size}. Our evaluation attested the entire program, unlike OAT and C-FLAT’s partial attestation or Blast’s single-path evaluation.
For the \texttt{move-syringe path} (bolus sizes 0.10, 0.50, 1, and 2 ml), \enola achieved the lowest average execution time overhead of \textbf{1.3}\% and compared to Blast’s 6.6\% at function-level granularity (Figure~\ref{fig:syring}).
\enola’s maximum $Auth$ size was \textbf{88} bytes, far smaller than C-FLAT’s 1,179 bytes for partial attestation; OAT’s size was smaller (69 bytes) but excluded loops.
In the \texttt{set-quantity path}, \enola incurred a 67.9\% overhead with a \textbf{48} bytes $Auth$ due to the small loop-based character-to-integer conversion, whereas C-FLAT’s exponential growth produced up to 1,527 bytes for larger inputs.

\textbf{Case study: anomalous behavior in the move-syringe path}.
We analyzed anomalous control-flow behavior in the \emph{move-syringe} path, similar to Blast, to demonstrate detection capability of \enola. The anomalous path, shown in Listing~\ref{code:syringe}, involves a loop iteration that dispenses or withdraws a specified bolus amount. The total motor steps are governed by the \texttt{steps} variable, which depends on \texttt{mLBolus} and \texttt{ustepsPerML} (Line 1). Here, \texttt{mLBolus} is the user input known to the \enola verifier, and the loop iteration count (\texttt{steps}) can be statically determined since \texttt{ustepsPerML} is constant. For example, bolus amounts of \texttt{0.010} and \texttt{0.011} correspond to 68 and 75 iterations, respectively.
As \enola instruments the beginning of loop bodies, 
the verifier - using offline program analysis and user input (\texttt{mLBolus}), can detect deviations from expected counts and flag anomalies.
\vspace{-0.3cm}
\begin{listing}[h]
	{\renewcommand\fcolorbox[4][]{\textcolor{gray}{\strut#4}}
		\begin{minted}[xleftmargin=5.8pt,numbersep=1pt, tabsize=2, frame=lines, framesep=1mm, breaklines,highlightlines={3}, linenos=true, escapeinside=||, fontsize=\scriptsize]{C}
			|\textbf{steps}| = mLBolus * ustepsPerML;
			for (i = 0; i < |\textbf{steps} |; i++){
				if(serialStr[0] == '+')
				dispense();
				else if(serialStr[0] == '-')
				withdraw(); }
		\end{minted}
	} 
	\cprotect\caption{Syringe pump: light blue - loop instrumentation}
	\label{code:syringe}
	\vspace{-0.1cm}
\end{listing}
\vspace{-0.3cm}
\subsection{Evaluations on Embench Applications}

\begin{figure*}[t]
	\centering
	\begin{subfigure}[t]{\textwidth} 
		\centering
		\includegraphics[width =.95\textwidth]{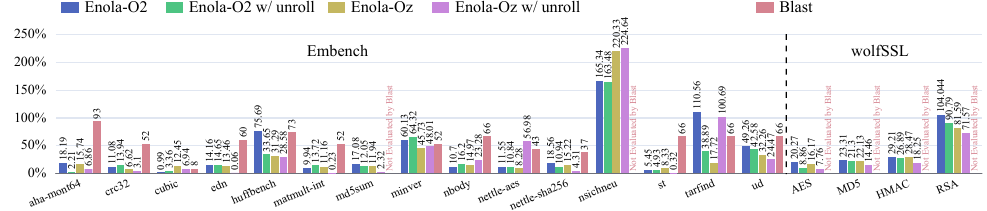}
		\vspace{-0.2cm}	
		\caption{Code size overhead (\%)} 
		\label{fig:bin_chart}
	\end{subfigure}
	\begin{subfigure}[t]{\textwidth} 
		\centering
		\includegraphics[width =.95\textwidth]{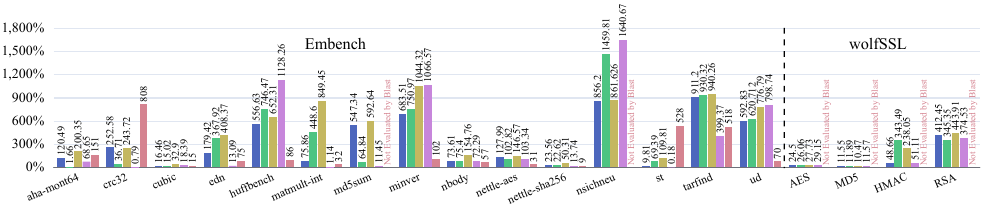}
		\vspace{-0.2cm}	
		\caption{Execution time overhead (\%)} 
		\label{fig:exe_chart}
	\end{subfigure}
	\begin{subfigure}[t]{\textwidth} 
		\centering
		\includegraphics[width =.95\textwidth]{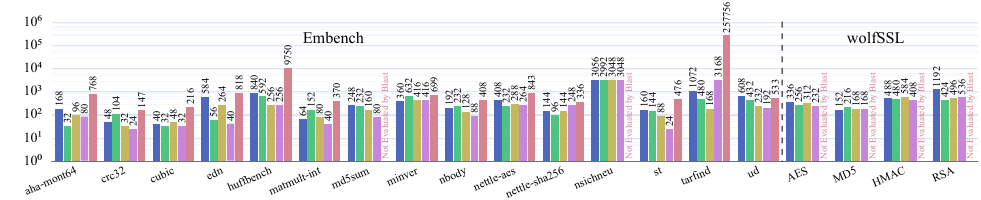}
		\vspace{-0.2cm}			
		\caption{$Auth$ size in bytes} 
		\label{fig:trace_chart}
	\end{subfigure}
	\vspace{-0.2cm}		
	\caption{Code size, $Auth$, and execution time overhead comparison between \enola and Blast}
	\label{fig:comparison}
	\vspace{-0.2cm}	
\end{figure*}

We evaluate Lines of Code (LoC) and CFG statistics of the Embench~\cite{Embench} applications, along with the \enola instrumentation sites.
We also evaluated Embench applications with loop unrolling by using the \texttt{-mllvm-unroll-count} compiler flag. 
Figure~\ref{fig:comparison} details the percentage increase with and without loop unrolling in code size due to \enola, execution time overhead, the resulting $Auth$ size in bytes, and comparisons of each with Blast.
We selected Blast for comparison because it demonstrates the best performance among previous works, and has also been evaluated using Embench.


\textbf{Code size overhead}.
With \texttt{O2}, \enola incurs \textbf{29.71}\% and \textbf{38.57}\% overhead with unrolling; under \texttt{Oz}, \textbf{35.3}\% and \textbf{32.62}\%, respectively.
Code size overhead depends on the number of instrumented CFG nodes by \enola.
Our results show that \texttt{Oz} induces fewer instrumentation sites in the CFG compared to \texttt{O2}, resulting in reduced code size overhead.
An exception is \texttt{nsichneu}, with similar instrumentation sites on both optimizations, but a smaller original code base amplifies the same instrumentation cost in \texttt{Oz}.
\enola overhead is significantly lower than Blast (64\%) on Embench, even without including complex applications like \texttt{nsichneu}.
This is due to fewer instrumentation sites for direct branches, function calls, address masking, and runtime path ID assignments.
OAT reported a 13\% size overhead, but it's evaluated on partial programs without loops or function calls.
\vspace{-0.1cm}		
\begin{figure}[h]	
	\includegraphics[width = .47\textwidth]{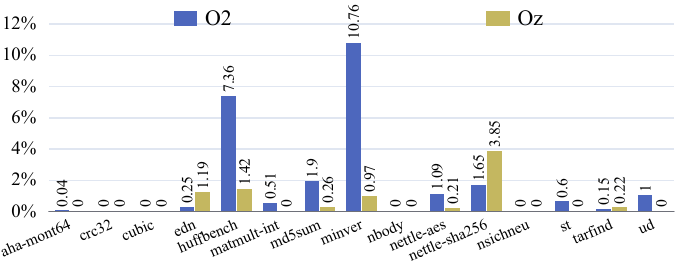}
	\vspace{-0.1cm}		
	\caption{Code size overhead due to register reservation} 
	\label{fig:reg-ovrhd}
	\vspace{-0.2cm}	
\end{figure}

We also assessed the impact on code size caused by reserving two measurement registers for \enola.
Figure~\ref{fig:reg-ovrhd} shows the effect of register reservations on application code size for both optimization levels.
On average, \texttt{O2} have a higher overhead, as it applies more aggressive optimizations compared to \texttt{Oz}.

\textbf{Execution time overhead}.
\enola incurs lower average execution overhead under \texttt{O2} than \texttt{Oz}.
Although \texttt{Oz} produces fewer instrumented basic blocks, code shrinking increases repeated executions of those blocks, resulting in higher runtime cost. 
For example, \texttt{st} shows 9.81\% overhead with \texttt{O2} (29 instrumentation sites), while \texttt{Oz} incurs 11$\times$ higher overhead despite only 13 sites, due to increased trampoline or instrumented code invocations (18,852 vs.\ 230,101).
In addition, compiler optimizations such as loop unrolling reduces the execution overhead of \enola as shown in Figure~\ref{fig:exe_chart}.


Compared with prior work, \enola demonstrates more realistic overheads, as earlier systems were evaluated on multicore Cortex-A CPUs with limited or coarse-grained attestation. 
Blast reports 185\% overhead with parallel log commit and 175\% in single-threaded (requiring function inlining), while providing only function-level verification; \enola supports fine-grained basic block-level verification. 
Even then, Figure~\ref{fig:exe_chart} illustrates that for six applications: \texttt{aha-mont64}, \texttt{crc32}, \texttt{edn}, \texttt{matmult-int}, \texttt{st}, and \texttt{tarfind}, \enola outperforms Blast on at least one optimization level.
Across common benchmarks, \enola’s average overhead is 322.20\% and 278.64\% for \texttt{O2} (with/without loop unrolling), and 284.54\% and 431.53\% for \texttt{Oz}. 
Other schemes report higher overheads: OAT achieves 2.7\% only on partial code (excluding loops or function calls) but 546\% on full programs, while C-FLAT reaches 1004\%. 
Performance could be further improved by adopting function-level attestation similar to Blast, at the cost of reduced security.

\begin{figure}[t]	
	\includegraphics[width = .48\textwidth]{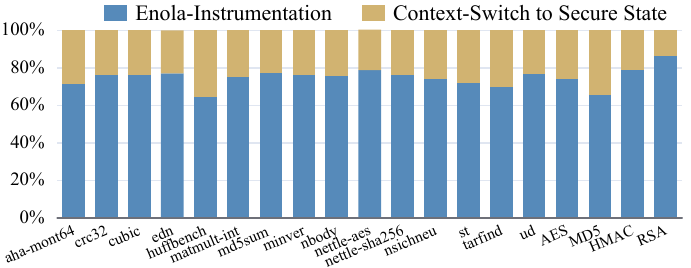}
	\caption{\enola runtime overhead breakdown for instrumentation (\texttt{O2}) and context-switch to secure state}
	\label{fig:ovrhd_breakdown}	
	\vspace{-0.4cm}
\end{figure}
\enola’s overhead arises from instrumentation (instrumented instructions and trampolines) and context switches to/from secure-state . 
As shown in Figure~\ref{fig:ovrhd_breakdown}, context switching accounts for about 26\% of total overhead under \texttt{O2}. 
This cost could be reduced by storing traces in normal state memory using SFI techniques as in Blast or utilizing unprivileged load/store instructions as in Silhouette~\cite{zhou2020silhouette}.

\textbf{$Auth$ size}.
Figure~\ref{fig:trace_chart} compares the $Auth$ sizes of \enola with four optimizations and Blast using Embench.
\enola with \texttt{Oz} produces fewer conditional branches compared to \texttt{O2}, leading to a reduced number of entries.
\enola $Auth$ depends on the number of unique $v_i.s$ encountered during program execution, while Blast requires separate entries for repeated execution of function calls or loops, inflating $Auth$ sizes.
For instance, \texttt{aha-mont64}, \texttt{huffbench}, and \texttt{tarfind} exhibit repeated execution of identical basic blocks within large or nested loops, leading to 768, 9,750, and 257,756 bytes $Auth$ size in Blast, compared to significantly smaller \textbf{168}, \textbf{840}, and \textbf{1,072} bytes in \enola under \texttt{O2}. 
An exception to this general trend is \texttt{ud} (\texttt{O2}), where Blast yields a smaller $Auth$ due to a higher frequency of conditional branches than function calls.
On average, \enola $Auth$ size under \texttt{O2} and \texttt{Oz} is \textbf{397} bytes and \textbf{224} bytes, respectively, while Blast produces an average of 21,009 bytes for the same set of applications, resulting in a \textbf{52} times or greater reduction in $Auth$ size.

\begin{figure}[t!]
	\centering
	\begin{subfigure}[t]{.49\columnwidth} 
		\centering
		\includegraphics[width = \linewidth]{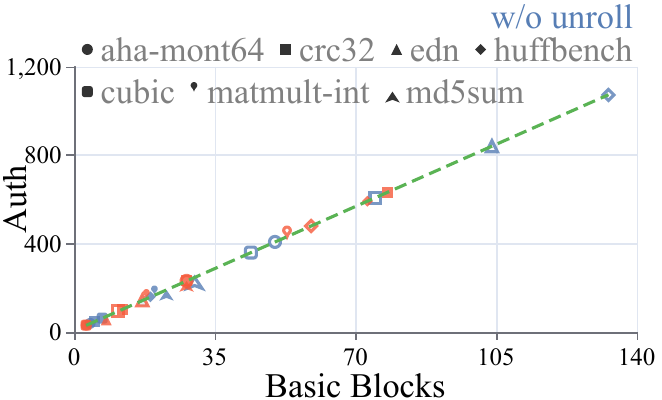}
		\label{fig:authVsBB}
	\end{subfigure}
	\begin{subfigure}[t]{.49\columnwidth} 
		\centering
		\includegraphics[width = \linewidth]{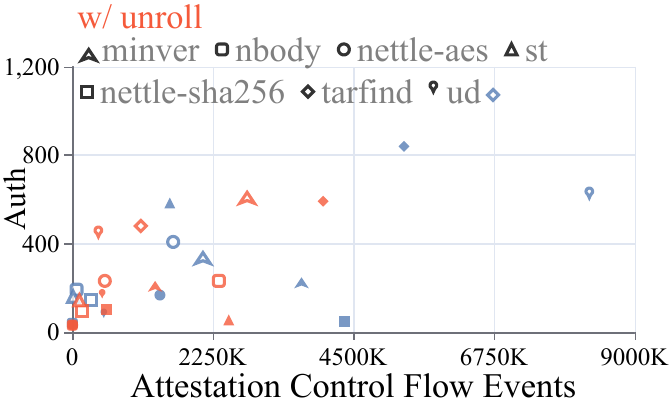}
		\label{fig:authVsCc}
	\end{subfigure}
	\vspace{-0.4cm}	
	\caption{\enola $Auth$ size linear to number of basic blocks instead of attestation control flow events}
	\label{fig:authVsBBvsCC}
	\vspace{-0.3cm}	
\end{figure}

To demonstrate that \enola generates $Auth$ with linear space complexity relative to the number of basic blocks, rather than control flow events as in existing works, we plot $Auth$ sizes for Embench in Figure~\ref{fig:authVsBBvsCC}.
The left side shows that the trace and measurement schemes scale linearly with basic blocks as discussed in Table~\ref{t:criteria} and Section~\ref{s:trace}.
While, the right side illustrates $Auth$ does not correlate with the number control flow events like existing approaches.
For instance, \texttt{aha-mont64} and \texttt{edn} exhibit a similar amount of control flow events (4,000K) but have 3 and 73 unique instrumented basic blocks, resulting in $Auth$ sizes of 32 and 592 bytes.

\subsection{Evaluations on wolfSSL Applications}

We evaluated four larger applications from wolfSSL~\cite{wolfSSL}, a library commonly used in embedded systems.
Specifically, we examined a 128-bit AES application processing 1KB of data, the MD5 message-digest, an HMAC application generating SHA256 hashes on 1KB of data, and a 2048-bit RSA asymmetric encryption application on 16 bytes of data.
Among them, RSA has the most complex CFG with 5,480 basic blocks and 3,309 of them were instrumented by \enola. 

\textbf{Code size and execution time overhead.}
Figures~\ref{fig:bin_chart} and~\ref{fig:exe_chart} also include the code size and execution time overheads on the wolfSSL applications. 
Although the average code size overhead closely resembles that of Embench, the average execution time is notably lower for both optimization levels.
This discrepancy can be attributed to the minimal overhead of the MD5 message digest application.
The loop unrolling optimization positively affects the code size and execution time overhead for large applications like RSA by reducing the conditional and loop condition basic blocks.

\textbf{$Auth$ size}.
Figure~\ref{fig:trace_chart} contains the $Auth$ sizes generated by \enola for the four wolfSSL applications.
They execute significantly more instrumented unique basic blocks compared to Embench, resulting in larger $Auth$ sizes.
Loop unrolling optimization for RSA (\texttt{O2}) reduced the unique basic blocks, thereby shrinking the $Auth$ size from \textbf{1,192} bytes to \textbf{424} bytes.
Overall, \enola $Auth$ scales linearly with the number of basic blocks even for larger applications.

Although \enola substantially reduces transmitted authenticator data, its runtime overhead is not uniformly low across all benchmarks.
The primary scalability benefit is communication-side: $Auth$ grows linearly with the number of basic blocks rather than the exponential trace length.
Runtime overhead depends on the frequency of instrumented branch targets, trace updates, and compiler optimizations; thus, tight loops or frequently executed instrumentions blocks can still incur noticeable overhead.
Therefore, \enola is best characterized as a communication-efficient and hardware-assisted CFA design.
Further runtime reductions are possible through coarser attestation granularity, as discussed in Section~\ref{s:trace}, fewer secure-state transitions, or hardware trace support.
\vspace{-0.2cm}
\subsection{Evaluations on Secure ISR Dispatcher}
We evaluated the performance of \enola interrupt dispatcher using a toggle-LED application.
The NVIC was configured to route the SysTick interrupt to the secure state, allowing the dispatcher to securely invoke the untrusted ISR in the normal state.
The SysTick timer generated an interrupt every 10ms, resulting in over 870.24 million interrupts during the application’s lifetime.
In the baseline configuration, the program finishes in 6.73 million clock cycles, whereas with the \enola dispatcher, it takes 7.53 million clock cycles, reflecting a modest overhead of \textbf{13.33}\% incurred by the runtime enforcement, MPU configuration, and context management.

%% file: related-work.tex
\section{Related Work}
\label{s:relatedWork}

\textbf{Attestation of embedded software integrity}.
Pioneer~\cite{seshadri2005pioneer} addresses the challenge of achieving verifiable code execution on untrusted legacy systems. The VIPER~\cite{li2011viper} approach extends these efforts by detecting peripheral firmware proxy attacks in remote attestation frameworks.
PUFatt~\cite{kong2014pufatt} combines physically unclonable functions with remote attestation to counter impersonation attacks. To systematize these advances, Armknecht et al.~\cite{armknecht2013security} proposed a mechanism for analyzing and designing software attestation schemes. In parallel, several works explored hardware-based remote attestation to establish a dynamic root of trust on untrusted platforms~\cite{eldefrawy2012smart}. While these static approaches can verify code integrity, they are limited in detecting dynamic control-flow hijacking attacks.

\textbf{Attestation of control-flow}.
Beyond C-FLAT, OAT, and Blast, several other control-flow attestation (CFA) approaches have been proposed. DIAT~\cite{abera2019diat} and ARI~\cite{wangari} employ software modularization based on criticality, combining monitoring and attestation of control and data flows. While those improve performance, they fail to detect sophisticated control-flow attacks. ReCFA~\cite{zhang2021recfa}, SpecCFA~\cite{caulfield2024speccfa}, and ScaRR~\cite{toffalini2019scarr} introduce coarse-grained program analysis to reduce the number of recorded entries through call-site filtering, control-flow event folding, and subpath checkpoint separation. ReCFA further leverages hardware memory protection for critical data structures and userspace kernel trapping, although these features are rarely available on embedded systems. ZEKRA~\cite{debes2023zekra} proposes a cryptographic method to delegate execution-path attestation to a dedicated third party, while subsequent works~\cite{ammar2025sok, caulfield2025run} analyze the capabilities of both provers and verifiers in CFA. CFA+~\cite{299850} leverages the ARMv8.5-A BTI feature to implement a hybrid mechanism enforcing local CFI and supports lightweight attestation, though without interrupt handling.

\textbf{Attestation via specialized hardware}.
VRASED, PEARTS, and TRACES employ modified/existing hardware for monitoring alongside software measurements to enable verification of program execution integrity~\cite{nunes2019vrased, joia2025provable}.
SANCUS utilizes customized hardware for static attestation~\cite{noorman2013sancus}.
ACFA~\cite{caulfield2023acfa}, LiteHAX~\cite{dessouky2018litehax}, Tiny-CFA~\cite{nunes2021tiny}, and LO-FAT~\cite{dessouky2017fat} are other approaches to CFA through specialized hardware, thus unsuitable for commodity devices.
IDA~\cite{arkannezhadida} leverages specialized hardware to monitor program counters, IRQs, memory addresses, and DMA writes, facilitating interrupt-aware attestation.
To tackle the TOCTOU challenge, RATA developed a specialized hardware aimed at reducing the time gap between the attestation state and reporting~\cite{de2021toctou}.
\enola is complementary to the TOCTOU defense approaches and focuses on full program attestation without hardware modification.

\textbf{ARM PA usage beyond authentication}.
Several recent works leveraged ARM PA for return address protection with measurement chains, providing spatial and temporal memory protection, pointer integrity, etc.
As a trusted measurement module on the normal state of TrustZone and digest storage on the higher bits of the pointer itself in Cortex-A, has led to efficient temporal and spatial memory protection in recent works of PTAuth~\cite{farkhani2021ptauth}, PAC it up~\cite{liljestrand2019pac}, and PACMem~\cite{li2022pacmem}.
PACStack overcomes the limitation of hash collision by constructing a measurement chain for all return addresses in the program~\cite{liljestrand2021pacstack}.
\enola is the first work to utilize PA for control-flow attestation on embedded systems.



%% file: bio.tex
\vspace{-0.5cm}
\begin{IEEEbiography}
	[{\includegraphics[width=1in,height=1.25in,clip,keepaspectratio]
		{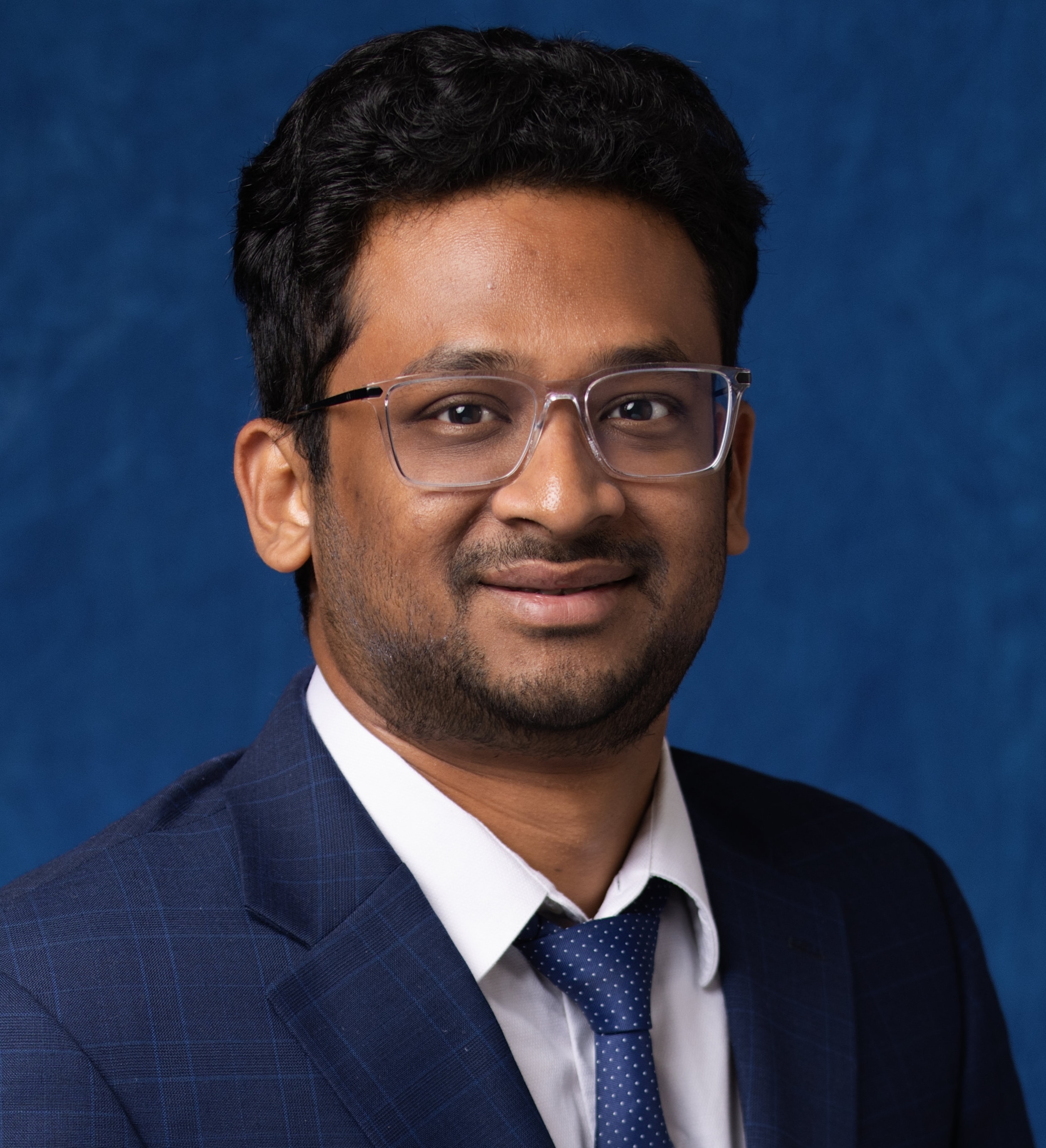}}]
	{Md Armanuzzaman}
	is an Assistant Professor in
	the Department of Computer Science at The University of Texas at El Paso,
	where he directs the Trusted Systems and Software Security Laboratory
	(T3SLab). His research interests include systems and software security for embedded systems, confidential computing, agentic systems, and program analysis. His work has appeared
	in venues such as ACM CCS, NDSS, ACM AsiaCCS, ACM EMSOFT, IEEE TIFS,
	IEEE SEED, and ACM SAC.
\end{IEEEbiography}

\begin{IEEEbiography}
	[{\includegraphics[width=1in,height=1.25in,clip,keepaspectratio]
		{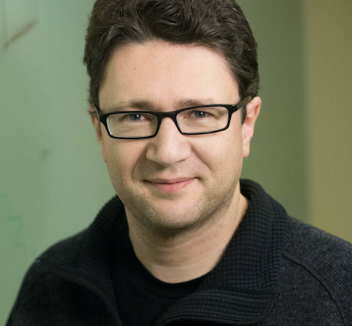}}]
	{Engin Kirda}
	 is a professor of computer science at Northeastern University in Boston. He previously held faculty positions at Institute Eurecom and the Technical University of Vienna, where he co-founded the Secure Systems Lab. His research focuses on malware analysis and detection, web application security, and social networking security, with more than 175 peer-reviewed publications. He has chaired top-tier security conferences, including NDSS 2015, USENIX Security 2017, AsiaCCS 2019, and ACM CCS 2023 and 2024. He co-founded Lastline, Inc., a malware-detection company acquired by VMware in 2020 and now part of Broadcom.
	\end{IEEEbiography}

\begin{IEEEbiography}
	[{\includegraphics[width=1in,height=1.25in,clip,keepaspectratio]
		{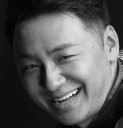}}]
	{Ziming Zhao}
	is an Associate Professor at the Khoury College of Computer Sciences and the director of the CyberspACe securiTy and forensIcs Lab (CactiLab), Northeastern University.
	His research has been supported by the U.S. National Science Foundation (NSF), the U.S. Department of Defense, the U.S. Air Force Office of Scientific Research, and the U.S. National Centers of Academic Excellence in Cybersecurity.
	He is a recipient of an NSF CAREER award and an NSF CRII award.
	He is also a recipient of best paper awards from USENIX Security 2019, ACM AsiaCCS 2022, ACM CODASPY 2014, and ITU Kaleidoscope 2016.
	\end{IEEEbiography}

%% file: apndx.tex
\appendix
\section*{ARMv8.1-M Pointer Authentication}
Table~\ref{t:pac-ins} represents the new instructions for ARMv8.1 pointer authentication (PA) security extension.
These PA instructions can generate and verify a keyed tweakable Pointer Authentication Code (PAC) for a pointer or data with the QARMA block cipher~\cite{avanzi2017qarma}.
The resulting 32-bit PAC is stored in a general-purpose register.
For example, the \texttt{pac} instruction signs the value in \texttt{lr} using \texttt{sp} as the tweak/- modifier and the key of the current state and privilege level and stores the result in \texttt{r12}.
With the \texttt{pacg} instruction, the software can specify which registers to use.
In the case of authentication failure, the authentication instructions, e.g., \texttt{aut}, generate an \texttt{INVSTATE} UsageFault.

\begin{table}[h]
	\centering
	\small
	\resizebox{\columnwidth}{!}{
		\begin{tabular}{l|l}
			\hline
			\multicolumn{1}{c|}{PA Instructions} & \multicolumn{1}{c}{Usage}\\ \hline
			\multirow{2}{*}{\texttt{pac}} & Sign \texttt{lr} using \texttt{sp} as the modifier and store \\ & the resulting PAC in \texttt{r12} \\ \hline
			\multirow{2}{*}{\texttt{aut}} & Authenticate \texttt{lr} using \texttt{sp} as the modifier \\ & and validate with the PAC in \texttt{r12} \\ \hline \hline	 	
			\multirow{2}{*}{\texttt{pacg} \texttt{rd}, \texttt{rn}, \texttt{rm}} & Sign a general-purpose register \texttt{rn} using \\ &\texttt{rm} as the modifier and store the PAC in \texttt{rd} \\ \hline
			\multirow{2}{*}{\texttt{autg} \texttt{rd}, \texttt{rn}, \texttt{rm}} & Authenticate a general-purpose register \texttt{rn} using\\ &\texttt{rm} as modifier and validate with the PAC in \texttt{rd} \\ \hline \hline
			\multirow{2}{*}{\texttt{bxaut} \texttt{rd}, \texttt{rn}, \texttt{rm}} & Authenticate \texttt{rn} using \texttt{rm} as the modifier \\&and \texttt{rd} as the PAC. If validated, branch to \texttt{rn}. \\ \hline
	\end{tabular}}
	\caption{The ARMv8.1-M PA extension instructions. \enola utilizes \texttt{pacg} for measurement calculations.}
	\label{t:pac-ins}
\end{table}

Table~\ref{t:pac-reg} displays the four 128-bit PA key registers, indicating their usage and configurable settings.
Each 128-bit PA key consists of four 32-bit registers, which are not memory mapped.
For example, \texttt{pac\_key\_u\_ns} register is made up of \texttt{pac\_key\_u\_ns\_0}, \texttt{pac\_key\_u\_ns\_1}, \texttt{pac\_key\_u\_ns\_2}, and \texttt{pac\_key\_u\_ns\_3} registers.
When a \texttt{pacg} instruction is executed at the unprivileged level and non-secure state, the \texttt{pac\_key\_u\_ns} register is implicitly used as the key to compute the PAC.
As denoted in the table, this PA key register can only be configured by software in the secure state and set individually for each 32-bit internal register using privileged \texttt{msr} (move-to-system-register) instructions.

\begin{table}[h!]
	\centering
	\small
	\resizebox{.70\columnwidth}{!}{
		\begin{threeparttable}
			\begin{tabular}{c|c|c}
				\hline
				PA Key Registers &  Used at &  Configurable at \\ \hline
				\texttt{pac\_key\_u\_ns} & U-NS & P-NS and P-S \\ \hline			
				\texttt{pac\_key\_p\_ns} & P-NS & P-NS and P-S \\ \hline
				\texttt{pac\_key\_u\_s} & U-S & P-S \\ \hline			
				\texttt{pac\_key\_p\_s} & P-S & P-S \\ \hline
			\end{tabular}
			\begin{footnotesize}
				U-NS: unprivileged non-secure, P-NS: privileged non-secure, U-S: unprivileged secure, P-S: privileged secure. 
			\end{footnotesize}
	\end{threeparttable}}
	\caption{The 128-bit PA key registers (not memory mapped) and their usage and configuration settings.}
	\label{t:pac-reg}
\end{table}

\section*{Trampoline and \enola Attestation Engine functions}

Listing~\ref{code:secT} illustrates an \enola trampoline cross-state call that requires two additional instructions in the non-secure callable state compared to a normal function call.
While Listing~\ref{code:secT2} presents first instruction in the secure state trampoline of attestation engine to retentive the branch destination $v_i.s$ to \texttt{r0} general purpose register.
\begin{listing}[h]
	{\renewcommand\fcolorbox[4][]{\textcolor{gray}{\strut#4}}
		\begin{minted}[xleftmargin=5.8pt,numbersep=1pt, tabsize=2, framesep=1mm, frame=lines,breaklines, escapeinside=||, fontsize=\scriptsize]{tasm}
;Non-secure callable memory region
report_direct:	
0x101FFE00	sg							
0x101FFE04	b report_direct <0x100031D4>
		\end{minted}
	} 
	\cprotect\caption{Non-secure callable region trampoline for reporting direct branches}
	\label{code:secT}
\end{listing}

\begin{listing}[h!]
	{\renewcommand\fcolorbox[4][]{\textcolor{gray}{\strut#4}}
		\begin{minted}[xleftmargin=5.8pt,numbersep=1pt, tabsize=2, framesep=1mm, frame=lines,breaklines, escapeinside=||, fontsize=\scriptsize]{tasm}
;Secure memory region
report_direct:
0x100031D4	mov r0, lr
0x100031D8	...
		\end{minted}
	} 
	\cprotect\caption{\enola attestation engine retrieving attested program's direct branch destinations}
	\label{code:secT2}
\end{listing}

\section*{Instrumentation for Leaf function return}
For leaf functions, the return address is stored in the \texttt{lr} register, allowing straightforward instrumentation to construct the backward edge measurement chain $M_b$ with a single instruction: \texttt{"pacg r11, lr, r11"}, as shown in Listing~\ref{code:lret}.
Here, the leaf and non-leaf function returns are chained together to construct $M_b$ on the reserved \texttt{r11} register.

\begin{listing}[t]
	{\renewcommand\fcolorbox[4][]{\textcolor{gray}{\strut#4}}
		\begin{minted}[xleftmargin=5.8pt,numbersep=1pt, tabsize=2, frame=lines, framesep=1mm, breaklines, highlightlines={3}, linenos=true, escapeinside=||, fontsize=\scriptsize]{tasm}
		epilogue:
			add	sp, #0x8
			pacg	r11, lr, r11 ;update measurement |$M_b$|
			bx lr
		\end{minted}
	} 
	\cprotect\caption{Instrumentation example: leaf function return}
	\label{code:lret}
\end{listing}

\section*{Verification Algorithm Details}
\label{appndx:verfi}


The \enola verifier uses the backtracking algorithm listed in Algorithm~\ref{algo_verify} to verify the legitimacy of the attested control path.
The algorithm takes the following inputs: the attestation report $\mathcal{R}$ and the control-flow graph $G_\mathcal{P}$ of the attested program, incorporating the entry basic block ($\mathcal{P}_{entry}$) and the potential exit points ($\mathcal{P}_{exits}$).


The algorithm initially confirms the validity of the attestation report signature $Sig_{K_a}(Auth, c)$ (Line 2-3) and checks for any illegal indirect branch targets (Line 4-5).
Upon successful verification, the algorithm abstractly executes the program based on its CFG and verifies the forward and backward measurements.
The abstract execution starts at $\mathcal{P}_{entry}$ and recursively executes various branches.
The abstract execution initializes an empty simulated call stack ($S$) to keep track of return addresses. 

The algorithm executes one of four actions based on the type of the last instruction at address $v_{c}.e$ within the current basic block ($v_{c}$): 
(1) if the last instruction in $v_{c}$ signifies a program exit, this marks the end of $\mathcal{P}$. 
Here, the computed measurements are compared with the received values in $Auth$ (Line 12);
(2) for a function call at $v_{c}.e$, the verifier pushes the address of the subsequent instruction onto the simulated function call stack $S$ and proceeds to abstractly execute the call target (Line 14-15);
(3) at $v_{c}.e$, in the case of a conditional branch, the algorithm explores all potential paths emanating from the current basic block (Line 17). 
And the occurrence count for each target in $T_\mathcal{O}$ is checked. If the occurrence count is greater than zero, the verifier decreases this count by one, computes a measurement ($M'_f$), and transitions abstract execution to that target. 
An invalid path is identified if no such targets exist in $T_\mathcal{O}$, prompting the recursive function to backtrack;
(4) when $v_{c}.e$ is a return instruction, the backward edge measurement ($M'_b$) is updated, and abstract execution moves to the return target from the simulated function call stack (Line 23-24).

\begin{algorithm}[ht!]
	\small
	\caption{\enola verifier}
	\label{algo_verify}
	\SetKwProg{Fn}{Procedure}{:}{}
	
	\KwIn{$G_\mathcal{P} = (V, E)$\\ 
		$\mathcal{R} = (Auth, Sig_{K_a}(Auth, c))$, where \\ $Auth =$ $(T_{\mathcal{O}}, \langle M_f, M_b \rangle)$ and \\ $T_{\mathcal{O}} = \{\langle v_i.s, \#v_i \rangle | i = 0,..., |V|-1\wedge\#v_i \neq 0\}, \{t_i| t_i \notin \bigcup_{i=0}^{|V|-1} v_i.s \}$}
	\KwOut{The legitimacy of the control-flow path}
	\BlankLine	
	\Fn{Verify($\mathcal{R}$)}{
		\If{$Sig_{K_a}(Auth, c)$ is not valid}
		{
			\KwRet{$False$}\;
		}
		\If{$|\{t_i\}| \not= 0$}
		{
			\KwRet{$False$}\;
		}
		\KwRet{AbstractExec($\mathcal{P}_{entry}, T_\mathcal{O}, 0, 0, \varnothing$)}\;
	}
	\BlankLine
	\Fn{AbstractExec($v, T, M_f', M_b', S$)}{
		$v_c \gets v$; $M_f'' \gets M_f'$;
		$M_b'' \gets M_b'$\;		
		$T' \gets T$; $S' \gets S$\;
		\If {$v_{c}.e \in \mathcal{P}_{exit}$}
		{
			$M_f'' \gets H_{K_m}(v_c.s, M_f')$\;
			\KwRet{$M_f == M_f'' \wedge M_b == M_b''$}\;
		}
		\ElseIf {$v_{c}.e$ is a function call}
		{
			$S'.push(v_{c}.e+1)$\;
			\KwRet{AbstractExec($e_{v_c}.d, T', M_f'', M_b'', S'$)}\;
		}
		\ElseIf {$v_c.e$ is not a function return}
		{
			\ForEach{ $d_i \in \cup e_{v_c}.d$}
			{
				\If {$T'.\#d_i$ > 0}
				{
					$T'.\#d_i \gets T'.\#d_i-1$\;
					$M_f'' = H_{K_m}(d_i, M_f')$\;
					\KwRet{AbstractExec($d_i, T', M_f'', M_b'', S'$)}\;
				}
			}
		}
		\ElseIf{$v_c.e$ is a function return}
		{
			$M_b'' = H_{K_m}(S'.top(), M_b') $\;
			\KwRet{AbstractExec($S'.pop(), T', M_f'', M_b'', S'$)}\;
		}		
		\KwRet $False$ \;
	}
\end{algorithm}

\section*{\texttt{pacg} versus software-implemented hash functions}
\begin{figure}[h!]	
	\includegraphics[width = .48\textwidth]{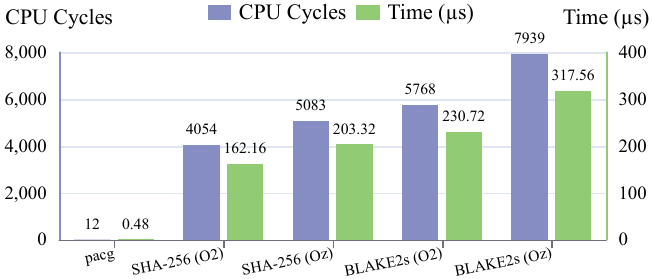}
	\caption{Execution time comparison: a single execution of the \texttt{pacg} instruction versus single runs of software-implemented hashing functions on a 25MHz Cortex-M85}
	\label{fig:pacg}
	\vspace{-0.3cm}	
\end{figure}

Figure~\ref{fig:pacg} presents a micro-performance analysis comparing a single execution of the \texttt{pacg} instruction for measurement with software implementations such as SHA-256 and BLAKE2s.
We evaluated the software measurement implementations at both \texttt{Oz} (optimized for size) and \texttt{O2} (optimized for speed) optimization levels.
As shown in the table, a \texttt{pacg} instruction consumes 12 CPU cycles, with \texttt{Oz}-optimized SHA-256 and BLAKE2s needing 5,083 and 7,939 cycles, respectively. 
In contrast, \texttt{O2}-optimized SHA-256 and BLAKE2s consume 4,054 and 5,768 cycles, respectively.
Given BLAKE2s' prevalence as a hash function in prior CFA solutions, a measurement calculation in \enola is at least 480 times faster than software-based measurements in earlier approaches.

\section*{\enola Micro-level Performance Evaluations}
Initializing PA key registers consumes approximately 112 CPU cycles, whereas a single instrumented direct branch, loop body, loop exit, and indirect call takes around 67, 71, 69, and 64 CPU cycles respectively, as shown in Table~\ref{t:break}.
The \enola attestation engine utilizes a table indexed by the starting address of each basic block ($v_i.s$) to store each basic block's occurrence trace. 
This structure allows for $O(1)$ time complexity when accessing and updating the occurrence trace.

\begin{table}[H]
	\centering
	\small
	\resizebox{.75\columnwidth}{!}{
		\begin{threeparttable}
			\begin{tabular}{c|c|c}
				\hline
				&  CPU cycles &  Time (\si{\micro\second}) \\ \hline
				Init. PA key registers & 112 & 4.48 \\ \hline			
				Instrumented direct branch &67 &2.79 \\ \hline
				Instrumented loop body &71 &2.84 \\ \hline
				Instrumented loop exit &69 &2.76 \\ \hline	
				Instrumented indirect call &64 &2.56 \\ \hline	 
			\end{tabular}
	\end{threeparttable}}
	\caption{\enola micro-level runtime overhead}
	\label{t:break}
	\vspace{-0.3cm}
\end{table}

Table~\ref{t:break} provides the micro-level runtime overheads for various \enola operations.
The \enola attestation engine initializes the PA key in approximately 4.8 \si{\micro\second}.
Runtime overheads for instrumentation at specific sites are as follows: 2.79 \si{\micro\second} for direct branches, 2.84 \si{\micro\second} for loop entries, 2.76 \si{\micro\second} for loop exits, and 2.56 \si{\micro\second} for indirect calls.

\section*{Evaluated application's CFG and \enola instrumentation sites}
Table~\ref{t:codeSize} and Table~\ref{t:LocO03} summarizes the Lines of Code (LoC), CFG statistics, and \enola instrumentation sites for the Embench and wolfSSL applications at \texttt{O0}, \texttt{O2}, \texttt{O3} and \texttt{Oz} optimization levels. Unlike prior control-flow attestation studies, \enola was evaluated on comparatively larger applications, notably the wolfSSL applications, using real-world compiler optimization levels.
\renewcommand{\arraystretch}{1}
\begin{table}[t]
	\centering
	\setlength\tabcolsep{.1ex}
	\small
	\resizebox{\columnwidth}{!}{
		\begin{tabular}{l|r|r|r|r|r||r|r|r|r|r|r}
			\hline
			\multirow{3}{*}{Application} & \multirow{3}{*}{LoC} &\multicolumn{4}{c||} {CFG Statistics}&\multicolumn{6}{c}{\enola Instrumentation Sites} \\ \cline{3-6} \cline{7-12}
			&  &  \multicolumn{2}{c|} {Nodes} &\multicolumn{2}{c||} {Edges}&\multicolumn{2}{c|}{Direct}&\multicolumn{2}{c|}{Indirect}&\multicolumn{2}{c}{Returns} \\ \cline{3-12}
			& &\multicolumn{1}{c|}{\texttt{O2}} &\multicolumn{1}{c|}{\texttt{Oz}} &\multicolumn{1}{c|}{\texttt{O2}} &\multicolumn{1}{c||}{\texttt{Oz}}&\multicolumn{1}{c|}{\texttt{O2}} &\multicolumn{1}{c|}{\texttt{Oz}} &\multicolumn{1}{c|}{\texttt{O2}} &\multicolumn{1}{c|}{\texttt{Oz}} & \multicolumn{1}{c|}{\texttt{O2}} &\multicolumn{1}{c}{\texttt{Oz}}  \\ \hline  \hline
			\multicolumn{12}{c}{Embench Applications} \\  \hline
			aha-mont64&162 &875 &898 &1,399 &1,422&27 &8 &0 &10&9&9\\ \hline			
			crc32&291 &385&414&289&303&25 &14 &0 &0&17&13\\ \hline
			cubic&254 &431&474&516&533&11 &21 &0 &11&6& 6\\ \hline
			edn &359&401&415&401&331& 47&28 &0 &0&13&13\\ \hline
			huffbench &309&653&481&793&455&230 &71 &0 &0&16&15\\ \hline
			matmult-int &175&369&406&288&317&23 &20 &0 &0&10&10\\ \hline
			md5sum &153&409&415&343&330&41 &25 &0 &0&15&15\\ \hline 
			minver &187&517&466&580&432&133 &58 &0 &0&7&7\\ \hline
			nbody &172&554&444&824&424&35 &23 &0 &0&7&7\\ \hline
			nettle-aes &1,147&440&461&412&398&73 &50 &0 &0&14&14\\ \hline
			nettle-sha256 &422&437&435&395&377&51 &30 &0 &0&10&10\\ \hline
			nsichneu&2,676&1,102&1,262&1,765&2,170&769&765 &0 &0&6&5\\ \hline
			st&117&856&438&1,723&404&29 &13 &0 &0&13&12\\ \hline
			tarfind&81&590&417&684&337&229 &28 &0 &0&16&13\\ \hline
			ud&95&435&403&417&324&82 &28 &0&0&6& 6\\ \hline \hline
			
			\multicolumn{12}{c}{wolfSSL Applications} \\  \hline
			AES&10,646& 1,038&1,087 & 1,648&1,702&90&87&0&0&23 &26\\ \hline
			MD5&4,917&1,194 &1,241 &2,002 &1,981&196&169&0&0&88 &70\\ \hline
			HMAC&6,085 &1,374&1,426&24,25&23,83&294&257&0&0&108&83\\ \hline
			RSA&27,807&5,480&3,444&9,813&6,110&3,309&1,476&0&18&272&267 \\ \hline
			
		\end{tabular}
	}
	\caption{\enola instrumentation sites, Lines of Code (LoC) and application's CFG statistics (\texttt{O2} \& \texttt{Oz}).}
	\label{t:codeSize}
\end{table}

\begin{table}[t]
	\centering
	\setlength\tabcolsep{.1ex}
	\small
	\resizebox{\columnwidth}{!}{
		\begin{tabular}{l|r|r|r|r|r||r|r|r|r|r|r}
			\hline
			\multirow{3}{*}{Application} & \multirow{3}{*}{LoC} &\multicolumn{4}{c||} {CFG Statistics}&\multicolumn{6}{c}{\enola Instrumentation Sites} \\ \cline{3-6} \cline{7-12}
			&  &  \multicolumn{2}{c|} {Nodes} &\multicolumn{2}{c||} {Edges}&\multicolumn{2}{c|}{Direct}&\multicolumn{2}{c|}{Indirect}&\multicolumn{2}{c}{Returns} \\ \cline{3-12}
			& &\multicolumn{1}{c|}{\texttt{O0}} &\multicolumn{1}{c|}{\texttt{O3}} &\multicolumn{1}{c|}{\texttt{O0}} &\multicolumn{1}{c||}{\texttt{O3}}&\multicolumn{1}{c|}{\texttt{O0}} &\multicolumn{1}{c|}{\texttt{O3}} &\multicolumn{1}{c|}{\texttt{O0}} &\multicolumn{1}{c|}{\texttt{O3}} & \multicolumn{1}{c|}{\texttt{O0}} &\multicolumn{1}{c}{\texttt{O3}}  \\ \hline  \hline
			\multicolumn{12}{c}{Embench Applications} \\  \hline
			aha-mont64&162 &146&98&199&121&16 &27&0 &0&9&9\\ \hline		
			crc32&291  &54&24&73 &33&16 &18&0& 0&8&10\\ \hline		
			cubic&254  &435&1,235&735 &2,355&16&9&0 &0&5&5\\ \hline		
			edn &359 &139&63&182&97&32&41&0&0&13&13\\ \hline		
			huffbench &309 &220&294&290&504&16&18&0&0&8&11\\ \hline		
			matmult-int &175 &70&58&94&76&16&11&0&0&10&10\\ \hline		
			md5sum &153 &80&75&119 &113&18 &18&0&0&11&11\\ \hline		
			minver &187 &198&266&260&462&58&133&0&0&8&7\\ \hline		
			nbody &172 &158&653&217&994&26&61&0&0&7&8\\ \hline		
			nettle-aes &1,147 &201&224&253&356&55&97&0&0&14&14\\ \hline		
			nettle-sha256 &422 &299&164&343&207&34&51&3&0&11&10\\ \hline		
			nsichneu&2,676 &1,581&782&2,232&1,546&897&769&0&0&5&6\\ \hline		
			st&117 &173&744&249&1,116&12&31&0&0&13&13\\ \hline		
			tarfind&81 &91&211&120&394&16 &18&0&0&8&11\\ \hline		
			ud&95 &100&216&129&377&30&88&0&0&6&6\\ \hline		\hline
			
			\multicolumn{12}{c}{wolfSSL Applications} \\  \hline
			AES&10,646 &306&103&413&165&148 &90&0 &0&27&25\\ \hline		
			MD5&4,917 &359&80&588&122&71&151&0&0&19&70\\ \hline		
			HMAC&6,085  &2,847&358& 4,683&524&71&109&0&0&19&109\\ \hline		
			RSA&27,807 &324&6,586&447&11,584& 2,404&12,079&0 &0&239&285\\ \hline		
			
		\end{tabular}
	}
	\caption{\enola instrumentation sites, Lines of Code (LoC) and application's CFG statistics (\texttt{O0} \& \texttt{O3}).}
	\label{t:LocO03}
\end{table}

\section*{Case Study on the crc32 Embench Application}
\label{appdx:crc32}

We conducted a case study on the Embench application performing a Cyclic Redundancy Check named \texttt{crc32} to illustrate the functionality of the \enola framework.
The source code of the program, depicted in Listing~\ref{code:crc}, reveals two main functions: the \texttt{benchmark\_body} parent function, which contains a loop that iteratively calls \texttt{crc32pseudo} based on the \texttt{rpt} value and the callee executes a loop 1,024 times to perform CRC operations.

\begin{listing}[t]
	{\renewcommand\fcolorbox[4][]{\textcolor{gray}{\strut#4}}
		\begin{minted}[xleftmargin=5.8pt,numbersep=1pt, tabsize=2, frame=lines, framesep=1mm, breaklines, linenos=true, highlightlines={5,7,14}, escapeinside=||, fontsize=\scriptsize]{C}
	DWORD crc32pseudo () {
		int i; register DWORD oldcrc32;
		oldcrc32 = 0xFFFFFFFF;
		for (i = 0; i < 1024; ++i) {//(2)crc loop-body branch
			oldcrc32 = UPDC32 (rand_beebs (), oldcrc32);
		}
		return ~oldcrc32;//(3)crc loop-exit branch
	}
	
	static int __attribute__ ((noinline))
	benchmark_body (int rpt) {
		int i; DWORD r;
		for (i = 0; i < rpt; i++) {//(1)App loop-body branch
			srand_beebs (0);
			r = crc32pseudo ();
		}
		return (int) (r % 32768); //App sequential return
	}
		\end{minted}
	} 
	\cprotect\caption{Source code of crc32 Embench application}
	\label{code:crc}
\end{listing}

\begin{figure}[h]
	\centering
	\includegraphics[width = .38\textwidth]{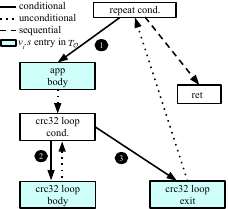}
	\caption{Simplified CFG of crc32 Embench application (\texttt{Oz})} 
	\label{fig:crcCFG}	
\end{figure}

For the \texttt{Oz} optimization level, the \enola compiler generates and instruments two basic blocks for loop body branches and one for loop exit branch, as highlighted in Listing~\ref{code:crc}.
Figure~\ref{fig:crcCFG} presents a simplified CFG of the program, where the entry address of highlighted nodes or basic blocks are designated as $v_i.s$.
Specifically, edges marked as \Circled{\textbf{1}} and \Circled{\textbf{2}} represent loop body control flow changes, while \Circled{\textbf{3}} denotes the loop exit edge in the \texttt{crc32pseudo} function.
Despite the presence of 303 edges in the CFG, only these three edges are executed at runtime that meet the criteria to be included in the trace.
Consequently, only the $v_i.s$ of the three highlighted basic blocks are included in the occurrence trace $T_\mathcal{O}$, resulting in a generated size of 24 bytes.
Table~\ref{t:crc32} illustrates the contents of the generated $T_\mathcal{O}$ when the application is executed with the \enola framework.
The $Auth$ consists of $T_\mathcal{O}$ combined with $M_f$ and $M_b$ measurements, which provide enough evidence for $\mathbb{V}$ to validate $\mathbb{P}$'s execution control-flow of the program $\mathcal{P}$ using \enola verification Algorithm~\ref{algo_verify}.

\begin{table}[H]
	\centering
	\resizebox{0.6\columnwidth}{!}{
	\begin{threeparttable}
		\begin{tabular}{c|c}
			\hline
			Address ($v_i.s$) &  Occurrence Count ($\#v_i$) \\ \hline
			0x10000495 & 4,250 \\ \hline
			0x10000441 & 4,352,000 \\ \hline
			0x10000465 & 4,250\\ \hline
		\end{tabular}
	\end{threeparttable}}
	\caption{$T_\mathcal{O}$ content of crc32 with \enola}
	\label{t:crc32}
	\vspace{-0.25cm}
\end{table}

\section*{Instruction Operands in $T_\mathcal{O}$}
Table~\ref{t:appA} shows the instructions meeting the criteria for \enola target basic block instrumentation and their corresponding elements or entries in $T_\mathcal{O}$.
The backward edges are enclosed in the measurement only, thus not included in $T_\mathcal{O}$.
\begin{table}[h]
	\centering
	\small
	\resizebox{0.8\columnwidth}{!}{
		\begin{threeparttable}
			\begin{tabular}{c|c|c}
				\hline
				Instruction Type &  Instruction &  $T_\mathcal{O}$ entry\\ \hline
				Indirect Call & \texttt{blx} rx & rx\\ \hline			
				Indirect Jump & \texttt{bx} rx & rx\\ \hline
				Conditional Jump & \texttt{b.\{condition\}} \#addr & \#addr \\ \hline
				Conditional Jump & \texttt{cb.\{condition\} \#addr} & \#addr \\ \hline
			\end{tabular}
	\end{threeparttable} }
	\caption{Instructions targeting \enola's instrumented basic blocks and corresponding $T_\mathcal{O}$ entry}
	\label{t:appA}
\end{table}

\section*{\enola Overhead for \texttt{O0} and \texttt{O3} Optimization Levels}
\label{s:O03}

\begin{figure*}[t]
	\centering
	\begin{subfigure}[t]{\textwidth} 
		\centering
		\includegraphics[width =.95\textwidth]{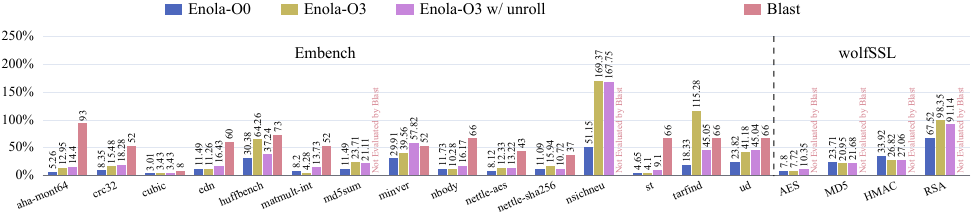}
		\vspace{-0.2cm}	
		\caption{Code size overhead (\%)} 
		\label{fig:O03bin_chart}
	\end{subfigure}
	\begin{subfigure}[t]{\textwidth} 
		\centering
		\includegraphics[width =.95\textwidth]{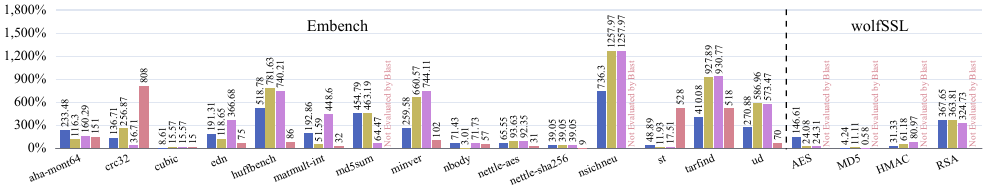}
		\vspace{-0.2cm}	
		\caption{Execution time overhead (\%)} 
		\label{fig:O03exe_chart}
	\end{subfigure}
	\begin{subfigure}[t]{\textwidth} 
		\centering
		\includegraphics[width =.95\textwidth]{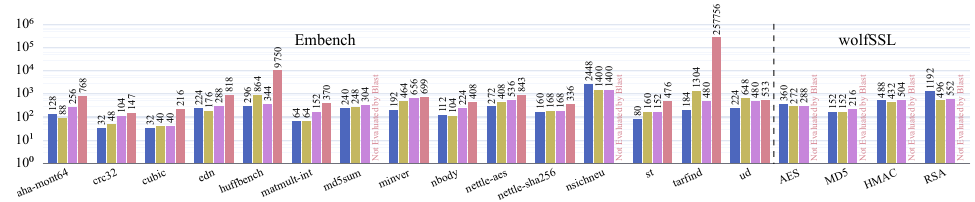}
		\vspace{-0.2cm}			
		\caption{$Auth$ size in bytes} 
		\label{fig:O03trace_chart}
	\end{subfigure}
	\vspace{-0.2cm}		
	\caption{Code size, $Auth$, and execution time overhead comparison between \enola and Blast}
	\label{fig:comparisonO03}
\end{figure*}

We further evaluate \enola under the \texttt{O0} and \texttt{O3} optimization levels, with and without loop unrolling, to provide a complete analysis across a broader spectrum of compiler behaviors.
At \texttt{O0}, the LLVM compiler disables all optimization passes, including loop transformations such as unrolling, resulting in larger and less efficient binaries.
Figure~\ref{fig:comparisonO03} presents the code size overhead, execution time overhead, and $Auth$ size for all 19 evaluated applications from Embench and wolfSSL.

As shown in Figure~\ref{fig:O03bin_chart}, \enola incurs an average code size overhead of \textbf{15.79}\% for Embench and \textbf{33.24}\% for wolfSSL applications under \texttt{O0}.
For \texttt{O3}, which applies more aggressive inlining and loop optimizations, the code size overhead increases to \textbf{36.23}\% with loop unrolling and \textbf{33.23}\% without for Embench.
In wolfSSL, the corresponding overheads are \textbf{37.71}\% and \textbf{38.46}\% respectively.
This trend is consistent with \texttt{O2}/\texttt{Oz} results, where aggressive optimizations tend to introduce more instruction reordering and duplication of measurement operations.
Compared to Blast, which adds 64\% overhead due to its extensive instrumentation of all control-flow events, \enola maintains a substantially lower footprint under all optimization levels by selectively instrumenting direct branches, function calls, and runtime path IDs.

At the \texttt{O0} level, \enola suffers from execution time overheads: 231.11\% for Embench and 137.46\% for wolfSSL, as shown in Figure~\ref{fig:O03exe_chart}.
For \texttt{O3}, \enola exhibits higher execution overheads of 358.98\% with loop unrolling and 370.63\% without for Embench.
For wolfSSL, the overhead is slightly lower: 115.04\% and 107.65\%, respectively.
These results indicate that although \texttt{O3} generates smaller baseline code, the instrumentation overhead becomes more prominent due to increased execution frequency of the optimized loops and functions.
Similar to \texttt{O2} and \texttt{Oz}, applying loop unrolling often results in fewer runtime invocations of the \enola trampoline logic.
However, for deeply nested or highly repetitive functions, unrolling can amplify the number of instrumented events, as observed in specific applications like \texttt{huffbench}.


As shown in Figure~\ref{fig:O03trace_chart}, \enola achieves a significantly smaller $Auth$ size compared to Blast.
This reduction stems from \enola’s design, which ensures linear trace generation by recording only unique control-flow transitions ($v_i.s$) during execution.
In contrast, Blast logs every direct function call and conditional branch instance, leading to a larger $Auth$ size, especially in applications with repetitive control paths.
The efficiency of \enola is evident when comparing the average $Auth$ size of the common Embench applications: \textbf{236} bytes under \texttt{O0} and \textbf{346} bytes under \texttt{O3}, in contrast to Blast, which generates an average of 21,009 bytes, resulting in nearly a \textbf{60x} reduction in transmission data.